\documentclass[11pt,a4paper]{article}
\usepackage[margin=1in]{geometry}
\usepackage{mathptmx,amsmath,amssymb,xfrac,mathtools,booktabs,longtable, multirow}
\usepackage{bbm}
\usepackage{lscape}
\usepackage[stable]{footmisc}
\usepackage{subcaption}
\usepackage{changepage}
\usepackage{textcomp,marvosym}
\usepackage{fixltx2e}
\usepackage{cite}
\usepackage{nameref,hyperref}
\usepackage[right]{lineno}
\usepackage{microtype}
\DisableLigatures[f]{encoding = *, family = * }
\usepackage[table]{xcolor}
\usepackage{array}
\usepackage[onehalfspacing]{setspace}

\newlength\savedwidth

\usepackage[aboveskip=1pt,labelfont=bf,labelsep=period,justification=raggedright,singlelinecheck=off]{caption}

\makeatletter
\renewcommand{\@biblabel}[1]{\quad[#1]}
\makeatother

\begin{document}

\begin{flushleft}
{\Large
\textbf{A decision integration strategy for short-term demand forecasting and ordering for red blood cell components}
}
\newline
\\
Na Li\textsuperscript{1,2*},
Fei Chiang\textsuperscript{1},
Douglas G. Down\textsuperscript{1},
Nancy M. Heddle\textsuperscript{2,3},
\\
\bigskip
\textbf{1} Department of Computing and Software, McMaster University, Hamilton, Ontario L8S 4L8, Canada
\\
\textbf{2} McMaster Centre for Transfusion Research, Department of Medicine, McMaster University, Hamilton, Ontario L8S 4L8, Canada
\\
\textbf{3} Centre for Innovation, Canadian Blood Services, Ottawa, Ontario K1G 4J5, Canada

\end{flushleft}

\section*{Abstract}

Blood transfusion is one of the most crucial and commonly administered therapeutics worldwide. The need for more accurate and efficient ways to manage blood demand and supply is an increasing concern. Building a technology-based, robust blood demand and supply chain that can achieve the goals of reducing ordering frequency, inventory level, wastage and shortage, while maintaining the safety of blood usage, is essential in modern healthcare systems. In this study, we summarize the key challenges in current demand and supply management for red blood cells (RBCs). We combine ideas from statistical time series modeling, machine learning, and operations research in developing an ordering decision strategy for RBCs, through integrating a hybrid demand forecasting model using clinical predictors and a data-driven multi-period inventory problem considering inventory and reorder constraints. We have applied the integrated ordering strategy to the blood inventory management system in Hamilton, Ontario using a large clinical database from 2008 to 2018. The proposed hybrid demand forecasting model provides robust and accurate predictions, and identifies important clinical predictors for short-term RBC demand forecasting. Compared with the actual historical data, our integrated ordering strategy reduces the inventory level by 40\% and decreases the ordering frequency by 60\%, with low incidence of shortages and wastage due to expiration. If implemented successfully, our proposed strategy can achieve significant cost savings for healthcare systems and blood suppliers. The proposed ordering strategy is generalizable to other blood products or even other perishable products.
\\

\noindent \textbf{Keywords}: Demand forecasting; inventory management; data-driven; blood demand and supply chain; red blood cell components.

\bigskip
\bigskip

\par\noindent\rule{\textwidth}{0.5pt}
Contact: Na Li. Email: lin18@mcmaster.ca \\
This work has been submitted to Operations Research for Health Care for publication in August 2020.

\newpage
%\linenumbers
\section{Introduction}
Blood transfusion is one of the most crucial and commonly administered therapeutics worldwide. The need for more accurate and efficient ways to manage blood demand and supply is an increasing concern in many countries, including Canada. Building a technology-based, robust blood product demand and supply system that can achieve the goals of reducing wastage and shortage, while maintaining the safety of the blood system, is essential in modern health care systems.

\begin{figure}[!ht]
\centering \makebox[\textwidth]{
\includegraphics[width=\textwidth]{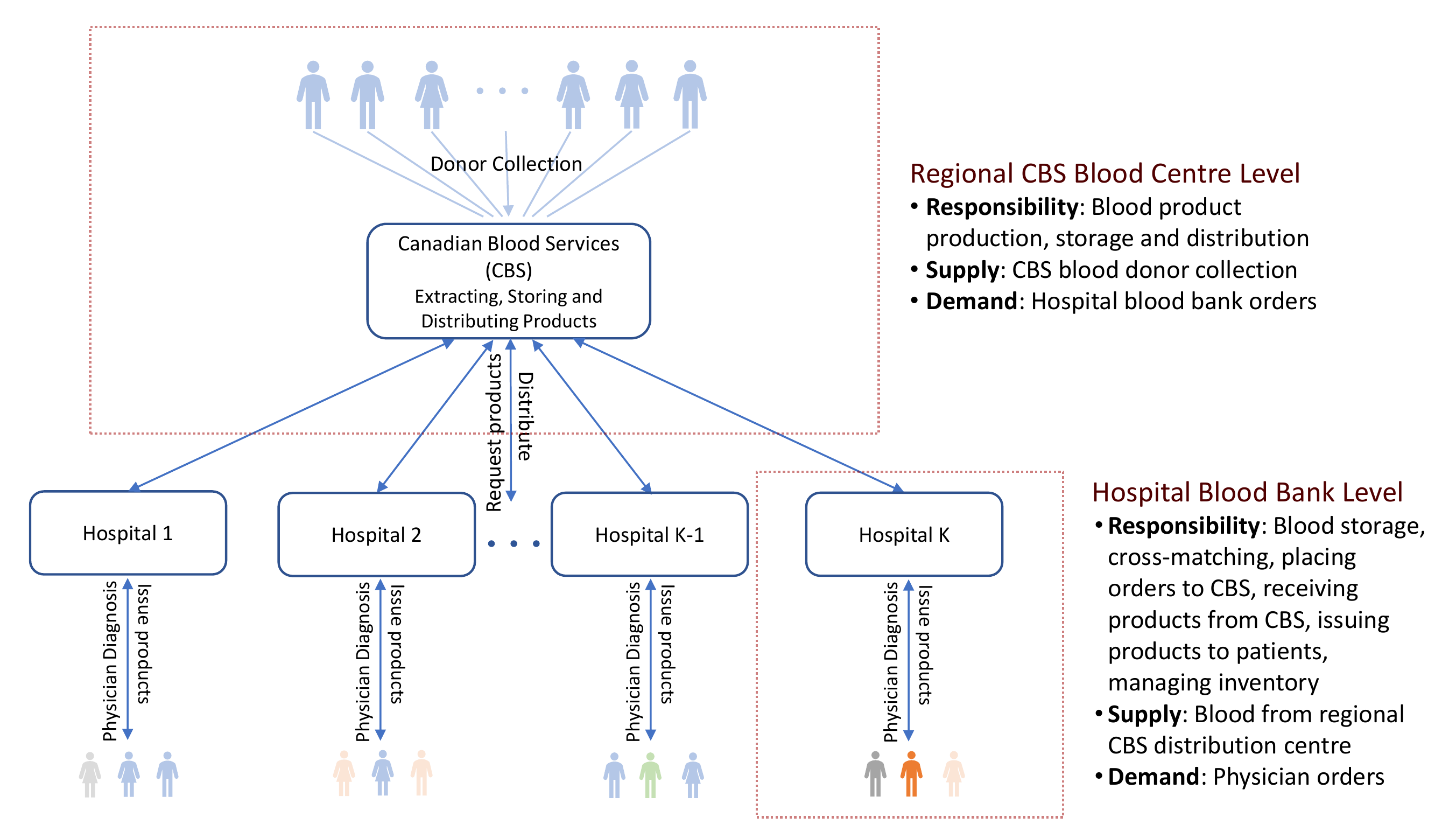}
}
\caption{Two-level supply chain with one regional blood centre and multiple hospitals}
\label{fig:flow}
\end{figure}

Canadian Blood Services (CBS) is the national blood supplier in Canada (excluding Qu\'{e}bec\footnote{Quebec has its own blood supplier, H\'{e}ma-Qu\'{e}bec, that supplies blood and other biological products of human origin to hospitals in the province.}). As shown in Figure \ref{fig:flow}, the current blood supply chain network in Canada is a centralized regional network consisting of two levels: regional CBS distribution centres and hospital blood banks. For example, there is one CBS regional blood distribution centre in Brampton, Ontario that covers the demands from the majority of hospitals in Ontario. There are nine regional CBS blood distribution centres across Canada \cite{cbscentre}. Each regional centre sets priorities to meet the demands of its own network; if there is excess supply, CBS decides centrally where the products are to be reallocated.

Currently, CBS has no information on recipient demographics and clinical utilization of the blood products distributed to hospitals. As a result, it has been very challenging for CBS to predict future demand and plan donor collection. For example, through this collaboration we discovered that decision makers at CBS have noticed a significant reduction in red blood cell (RBC) transfusions over recent years. They hypothesize this reduction could be caused by changes occurring in local hospitals, such as increased engagement on RBC clinical transfusion guidelines or improved surgical techniques. However, without clinical data, any such hypothesis cannot be investigated. In addition, the impact of clinical changes on the RBC demand cannot be evaluated. It is essential to maintain robust blood demand and supply management not only at the regional blood centre and individual hospital levels, but also as a complete supply chain. Increasing the transparency of blood utilization between the national blood supplier and the hospitals is important to promote efficient blood supply chain management (BSCM).

At the level of hospital blood banks, current inventory management practices rely heavily on human decisions. There are three key limitations at hospital blood banks: 1) The blood product ordering decisions are made from human experience. There is little quantitative evidence of the fluctuations in RBC demands to support decision making. 2) Since the electronic data systems in hospitals are primarily designed for clinical data collection, the systems do not have functions such as inventory reporting, forecasting or planning. 3) They are facing heterogeneous demand processes. Hospital demands are made by physicians based on individual patient diagnoses with different requirements, such as multiple units issued at the same time and units with specific dose or brand requirements. These limitations increase the complexity of inventory management. As a result, hospital blood banks tend to cope with the variation in demand by holding excess inventory.

While holding excess inventory can help hospital blood banks decrease the risk of shortages, it increases the holding costs for storing blood and the risk of wastage\footnote{Throughout this paper, the terms ``wastage" and ``waste" are all referring to the blood products wasted due to expiration.}. It may also lead to extra costs for reallocating units close to expiry. Moreover, studies \cite{Sparrow, Heddle_bloodage, JCI90837, Sut2017} have shown that duration of RBC storage can affect functional integrity and quality standards which could impact patient outcomes. Results of randomized trials suggest that there is no benefit to transfusing very fresh blood compared to blood stored for a longer duration (i.e. less than 10 days since blood donation compared with 24 days or more \cite{Middelburg2013}); however, it has been suggested that the clinical impact of blood storage over its 42-day shelf life may not be linear. More specifically, the risk of a specific outcome (i.e. mortality) by days of storage could be a convex curve \cite{Trivellal2320}. Through controlling inventory levels in hospital blood banks, it is possible to restrict the age of transfused blood into a reasonable range that may correlate with better patient outcomes.

Furthermore, from the national blood supplier's point of view, when hospital blood banks hold a large amount of inventory, it prevents CBS from understanding the real demand and restricts the ability of CBS to adjust for demand variability. As a result, both hospital blood banks and CBS cannot operate the demand and supply chain efficiently. With the availability of a large amount of clinical data and advanced analytical methodologies, there are opportunities to increase the accuracy of blood demand forecasting and improve the efficiency of the entire demand and supply chain.

Fresh blood components include RBC, platelets, and plasma. The RBC component experiences the highest demand among all the fresh components. It is prepared by removing plasma from a whole blood donation and is used to treat hemorrhages and to restore tissue oxygenation \cite{Klein2007}.  The demand for RBC components determines the plan for blood collection and production. In this study, we aim to tackle the demand forecasting and inventory management challenges for RBC components.

We study a large clinical database with over 1.2 million blood transfusions for nearly 100,000 patients in Hamilton, Ontario from 2008 to 2018. We perform a thorough investigation of RBC utilization with associated trends, and formulate a hybrid model for short-term RBC demand forecasting using clinical indicators. The demand forecasting model is then used to develop an integrated ordering strategy. The proposed integrated methodology achieves three goals: i) a more accurate forecasting method that reflects the actual RBC demand at hospital blood banks, which increases the transparency between CBS and hospital blood banks; ii) a leaner and fresher inventory at hospital blood banks, which may correlate with better patient outcomes; iii) a simpler ordering strategy that requires less frequent orders on scheduled days, which reduces human resources and costs both at hospitals blood banks and CBS. The main contributions of this study are as follows:
\begin{enumerate}
\item We summarize the key challenges in hospital blood banks and provide quantitative evidence to justify the need for a comprehensive methodology to resolve these issues.
\item We decompose the time series of daily RBC transfusions into trend, seasonality and residuals. We find different association patterns between the trend+seasonality and selected clinical indicators, versus between the residuals and selected clinical indicators, which motivates the development of a hybrid model for demand forecasting.
\item We develop and validate a hybrid model that combines Seasonal and Trend decomposition using Loess (STL) time series and eXtreme Gradient Boosting (XGBoost) models to forecast future RBC demands. The model is compared with an STL model alone, an STL + linear regression model, and a long short-term memory (LSTM) model. %Compared to the actual RBC demands in 2018, the mean absolute percentage error (MAPE) of the predicted daily demand using the hybrid model is 16.94\%, whereas the MAPE of the historical daily ordered quantities by hospital blood banks is 49.53\%.
\item We construct and solve a multi-period inventory problem, with constrained inventory target and reorder level, for an integrated ordering strategy using the demand forecasts from the proposed hybrid model. The cost function considers routine delivery cost, holding cost, same-day urgent delivery (shortage) cost, and wastage cost. We propose a daily ordering strategy, and a semiweekly ordering strategy that allows hospital blood banks to make orders twice a week, on Mondays and Thursdays. %The semiweekly ordering strategy achieves a reduction of 39\% in the inventory level and a reduction of 60.55\% in the ordering frequency, while no shortages or wastage are observed. It leads to a leaner inventory level, fresher blood, lower costs, and a simpler delivery schedule. It creates a ``win-win" situation from the perspectives of both CBS and hospital blood banks.
\end{enumerate}

In Section \ref{hospitalbloodbank}, we describe the inventory management process at hospital blood banks, and existing challenges. In Section \ref{literature}, we provide a literature review of demand forecasting models for blood products, methods for blood demand and supply chain management, and implementations for blood product management in healthcare systems. Section \ref{method} provides the data description, model background, model development, model training, model evaluation, and the proposed integrated ordering strategy for RBC inventory management. Section \ref{result} presents the results of the hybrid demand forecasting models for daily and semiweekly demand predictions, and the comparisons between the proposed ordering strategy and the existing strategy used in current hospital blood banks. Section \ref{discussion} discusses limitations and future opportunities of this study.

\section{Existing Challenges in Hospital Blood Banks}
\label{hospitalbloodbank}
This section describes the daily routine of blood product ordering and inventory management in a typical hospital blood bank. There are four hospital blood banks in Hamilton, Ontario operated by one Transfusion Medicine (TM) laboratory team consisting of four faculty members and three technical specialists spread across Hamilton General Hospital, Juravinski Hospital, McMaster University Medical Centre (MUMC)\footnote{The TM laboratory at MUMC is the hub site for performing all remote testing for the West Lincoln Memorial Hospital satellite site. MUMC is a children's hospital that has a small number of adult and pediatric outpatient clinics. This background description is based on the information collected at MUMC prior to the COVID-19 pandemic. Visits to other blood banks were cancelled for safety considerations during the pandemic situation. The hospital blood bank at MUMC has relatively smaller demands compared with other blood banks in Hamilton, Ontario. However, all hospital blood banks in Hamilton, Ontario are managed by the same Transfusion Medicine laboratory team.}, and St. Joseph's (STJ) Healthcare Hamilton. These laboratories manage and store all the blood products and fresh blood components in the Hamilton Region, as well as provide expert diagnostic, consultative and educational services for patients, physicians, and allied health care providers.

\noindent \textbf{Hospital blood bank daily routine:} The TM labs are responsible for ordering, storing, managing and issuing all fresh blood components including RBCs, platelet components, frozen plasma components, frozen cryoprecipitate and other experimental blood components such as COVID-19 convalescent plasma. At MUMC, every Monday to Saturday at approximately 4 pm, a technical specialist in the lab performs a routine physical count for each product in their inventory, compares the physical counts with the pre-defined inventory targets and sends an order to CBS (Brampton) through fax. The order delivery from the CBS distribution centre usually arrives between 8:30 am and 10 am on the next day. CBS provides a paper-copy summary list of the delivered products with the shipment. The technical specialist compares this summary list with the order form they faxed to ensure they received all the products they requested. (Feedback from the hospital blood bank indicates that only occasionally their orders are replaced, cancelled or delayed. However, due to the fact that there are no electronic data entries to trace the orders, a statistical summary of such information is not possible.) On the CBS side, a specific person manually transfers ordering information from faxed paper copies to an electronic spreadsheet\footnote{To our knowledge this ordering data collection process at CBS started in 2012.}. The data are collected for CBS administration and operation purposes and are not shared with hospital blood banks.

After a hospital blood bank receives blood products from CBS, the technical specialist enters the information into an electronic health system named Meditech\footnote{Meditech is the electronic data system used in Hamilton, Ontario. Other hospitals may use a different system. Even for the hospitals using Meditech, the design of the system could also be different.}. During the day, physicians may make prescription orders at any time through Meditech or by fax to the hospital blood banks. Usually, physicians make daily orders per patient, and the ordered products are available for pick up at different scheduled times. For every physician order, the technical specialist reviews the patient's historical clinical information which in this case is a custom Meditech report. The report includes the most recent lab test results and antibody screening results. These are used to determine whether special transfusion requirements are needed. Finally, a compatible product will be assigned to the patient.

\noindent \textbf{Current ordering strategy:} Using their experience, the technical specialists at each Hamilton hospital determine fixed inventory targets for different products. Orders are made to raise inventory up to these targets. We find the targets set by hospital blood banks are significantly higher than the actual demands. For example, Table \ref{tab:stock_rbc} shows the inventory targets for RBC units by blood group at MUMC. The total inventory target in the table is eight times larger than the mean daily demand at MUMC, which has the smallest number of days of inventory on hand among all hospital blood banks in Hamilton, Ontario. More details are described in Challenge 2 below. It is natural that hospital blood banks are most concerned about having sufficient inventory. However, without the quantitative evidence as we provide in this study, the consequence is that the order quantities cannot reflect fluctuations in the actual demand, resulting in excess inventory (prolonged days of inventory on hand), increased risk of wastage, and overly frequent same-day urgent orders.

\begin{table}[htbp]
\begin{tabular}{|l|p{0.1\textwidth}|>{\centering}p{0.1\textwidth}|>{\centering}p{0.1\textwidth}|>{\centering}p{0.1\textwidth}|>{\centering\arraybackslash}p{0.1\textwidth}|}
\hline
\multicolumn{2}{|l|}{\multirow{2}{0.25\textwidth}{\textbf{Number of units}}} & \multicolumn{4}{c|}{\textbf{Blood group}} \\ \cline{3-6}
\multicolumn{2}{|c|}{}  & O & A & B & AB\\ \hline
\multicolumn{1}{|c|}{\multirow{2}{0.15\textwidth}{\textbf{Rh type}}}  & Positive   & 20 & 20 & 8 & 0 \\ \cline{2-6}
\multicolumn{1}{|c|}{} & Negative & 12 & 8 & 2 & 0 \\ \hline
\end{tabular}%
\caption{\small{RBC inventory target by ABO Rh type in MUMC blood bank}}
\label{tab:stock_rbc}
\end{table}

\noindent \textbf{Challenge 1: Excess inventory level.} The mean and standard deviation (sd) statistics in Hamilton, Ontario for the days of inventory on hand, age (days) of blood, and daily number of units in-stock are shown in Table \ref{inventory_desc}. The mean (sd) of days of inventory on hand (DOH) in 2018 was 12.33 (8.62) days, and the mean (sd) of the age of blood prior to transfusion was 23.45 (10.58) days; whereas in 2008, the mean (sd) of DOH was 8.83 (6.82) days, and the mean (sd) of the age of blood prior to transfusion was 16.04 (8.41) days. The wastage rate for RBCs was slightly reduced from 2.14\% in 2008 to 1.68\% in 2018. From 2008 to 2018, there was a growth of 0.35 days per year for the mean DOH. It increased significantly between 2012 and 2015, then became more stable after 2016 (one-way ANOVA: $F=19.74, p=0.002$). There was a significant increase in the mean daily number of units in-stock over time (one-way ANOVA: $F=27.3, p<0.001$), i.e. on average 33.4 units per year. Similarly, there was a dramatic increase in the inventory level in 2012. Our proposed ordering strategy can efficiently reduce the mean and sd of the inventory level leading to a significant reduction in DOH.

\begin{table}[htbp]
\centering
\small
\begin{tabular}{|l|c|c|c|c|c|c|}
\hline
\textbf{Year} & \textbf{Days on hand (DOH)} & \textbf{Age (days) of blood} & \textbf{Number of units in-stock} & \textbf{Wastage (\%)} \\ \hline
2008 & 8.83 (6.82) & 16.04 (8.41) & 983.36 (118.96) & 2.14 \\ \hline
2009 & 9.70 (7.21) & 17.70 (8.29) & 1,035.33 (69.30) & 2.15 \\ \hline
2010 & 9.33 (7.46) & 19.07 (8.95) & 1,004.47 (57.20) & 2.06 \\ \hline
2011 & 9.32 (7.18) & 18.54 (8.26) & 1,101.58 (87.41) & 1.88 \\ \hline
2012 & 11.30 (8.45) & 20.80 (9.58) & 1,301.40 (107.56) & 1.97 \\ \hline
2013 & 12.33 (8.70) & 22.40 (10.05) & 1,358.97 (68.22) & 1.59 \\ \hline
2014 & 12.73 (8.94) & 22.01 (10.47) & 1,317.41 (68.64) & 1.48 \\ \hline
2015 & 13.07 (8.78) & 23.22 (10.66) & 1,328.80 (94.70) & 1.63 \\ \hline
2016 & 11.88 (8.31) & 19.23 (9.68) & 1,339.16 (75.35) & 1.45 \\ \hline
2017 & 11.87 (8.26) & 20.76 (9.58) & 1,376.59 (74.58) & 1.42 \\ \hline
2018 & 12.33 (8.62) & 23.45 (10.58) & 1,317.46 (65.89) & 1.68 \\ \hline
\end{tabular}
\caption{Mean (sd) of days of inventory on hand (DOH), age (days) of blood prior to transfusion, and daily number of units in-stock, as well as wastage rate by year in Hamilton hospital blood banks}
\label{inventory_desc}
\end{table}

\begin{figure}[htbp]
\centering \makebox[\textwidth]{
\includegraphics[width=0.5\textwidth]{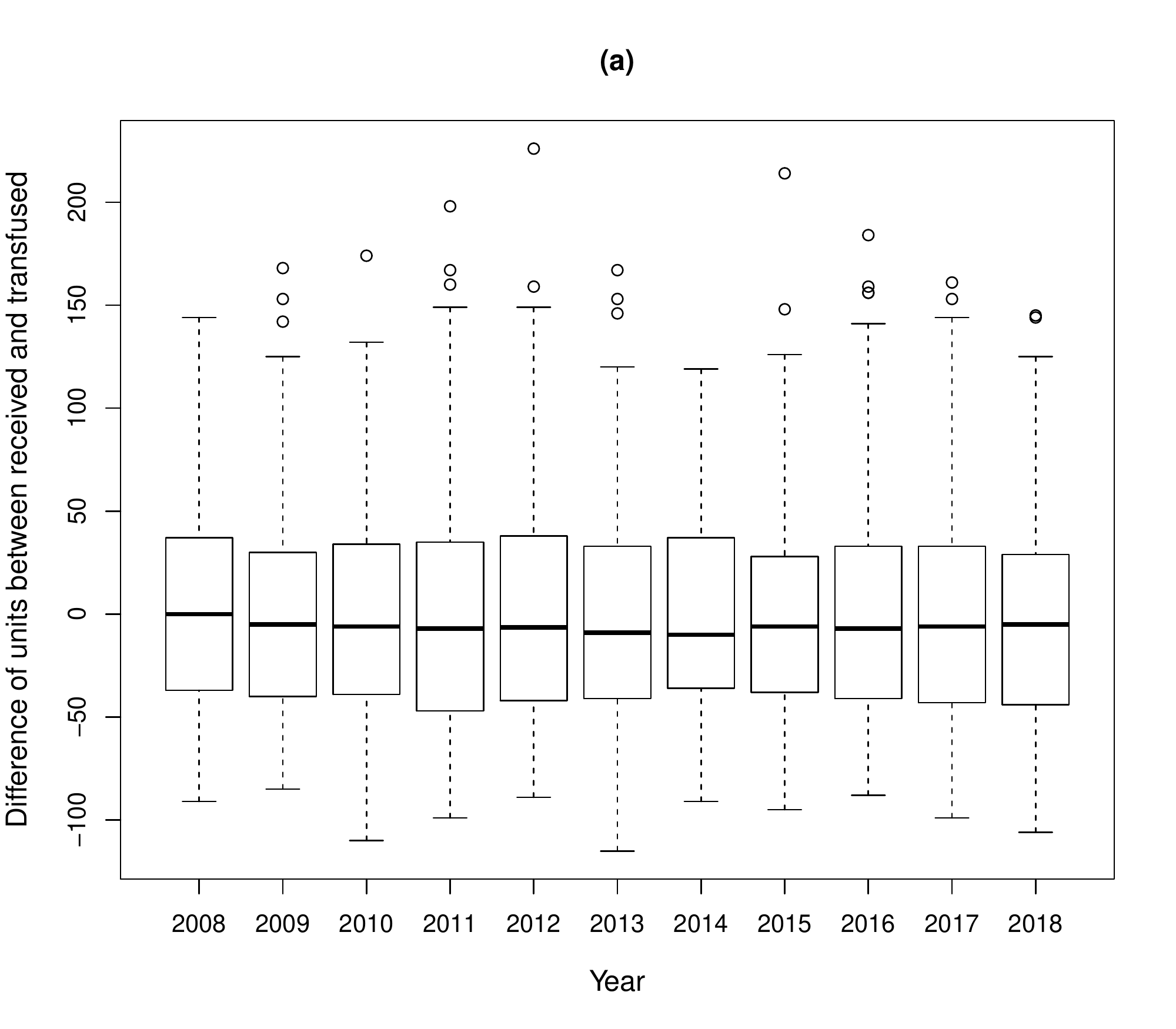}
\includegraphics[width=0.5\textwidth]{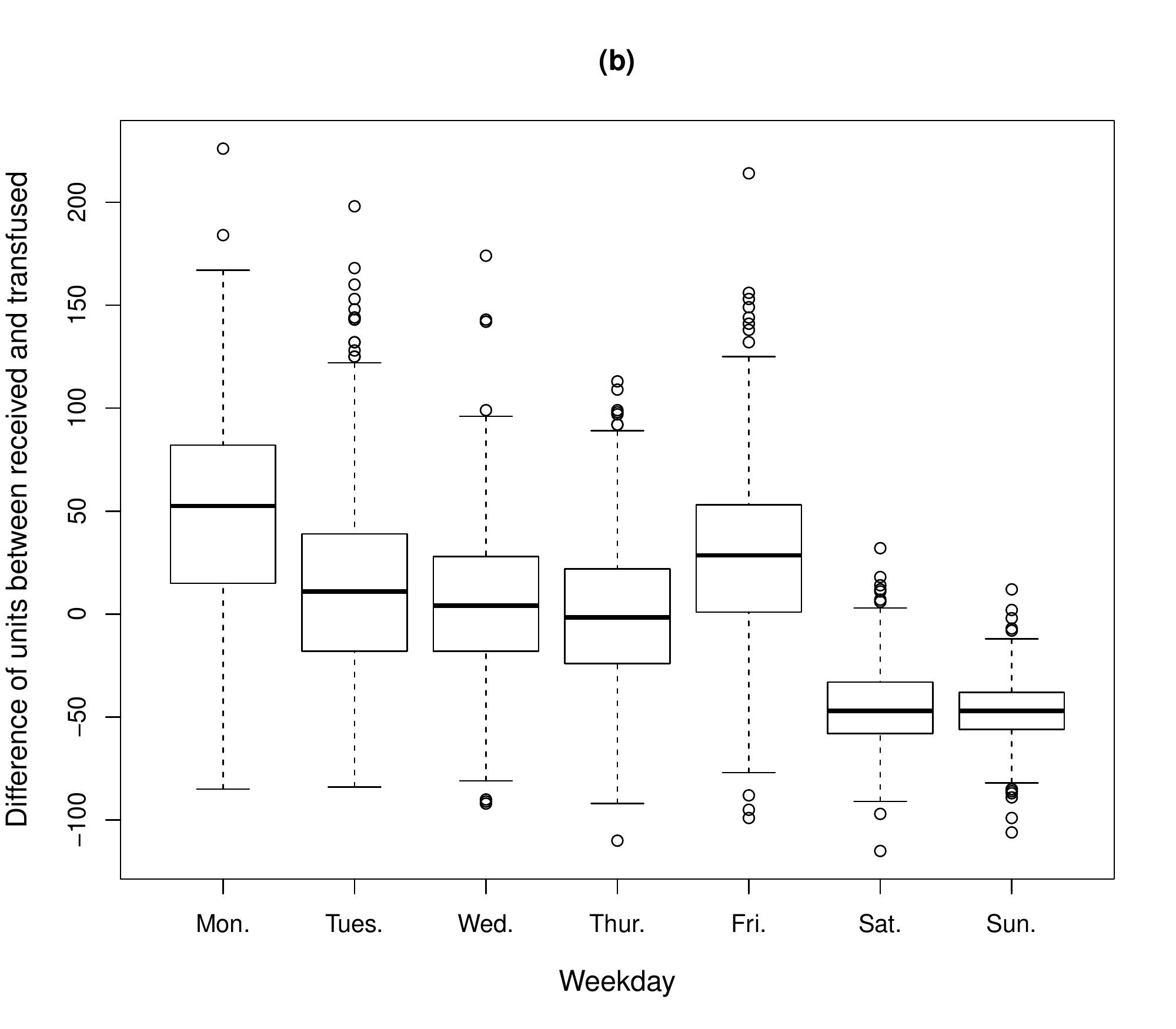}
}
\centering \makebox[\textwidth]{
\includegraphics[width=0.5\textwidth]{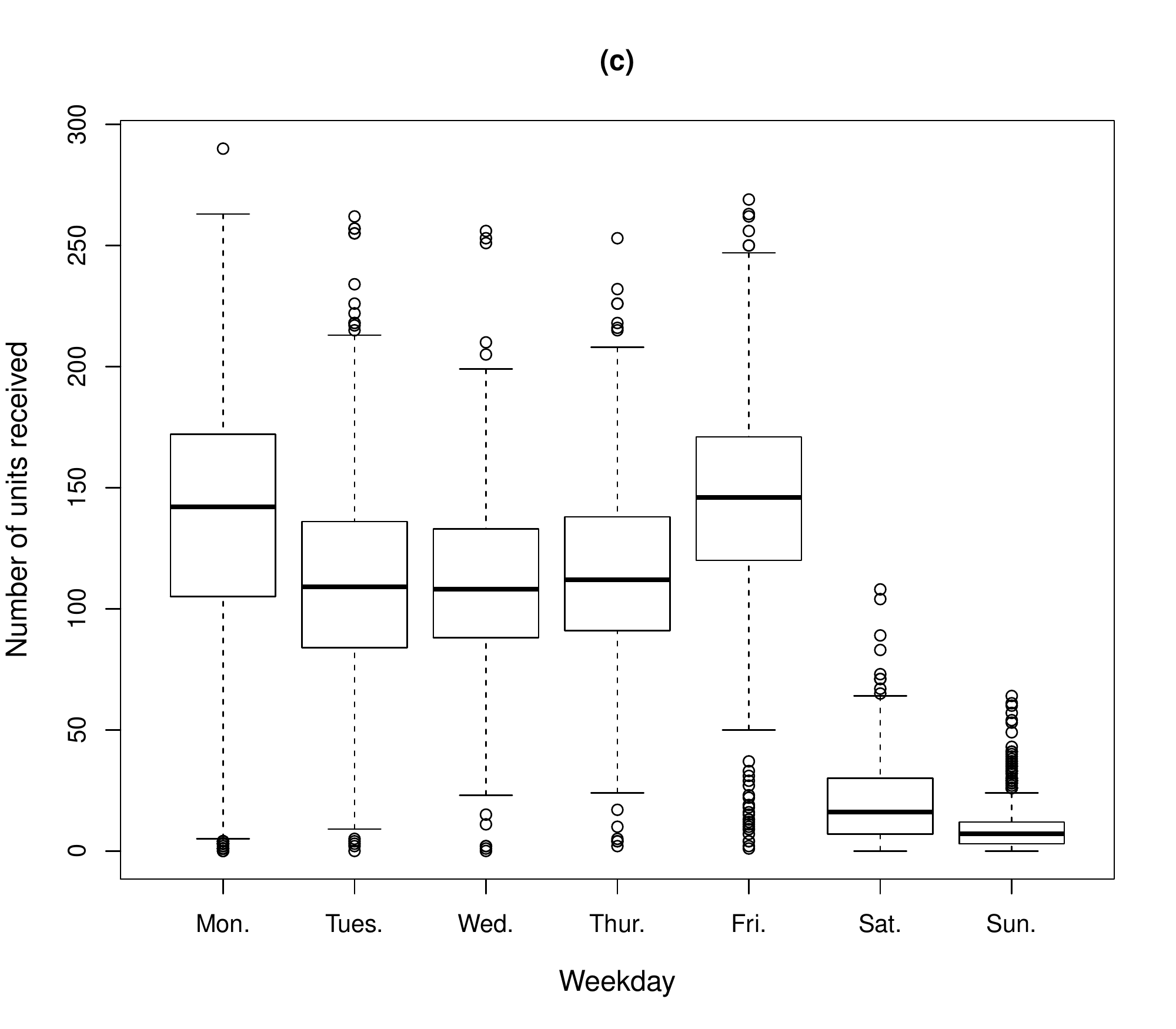}
\includegraphics[width=0.5\textwidth]{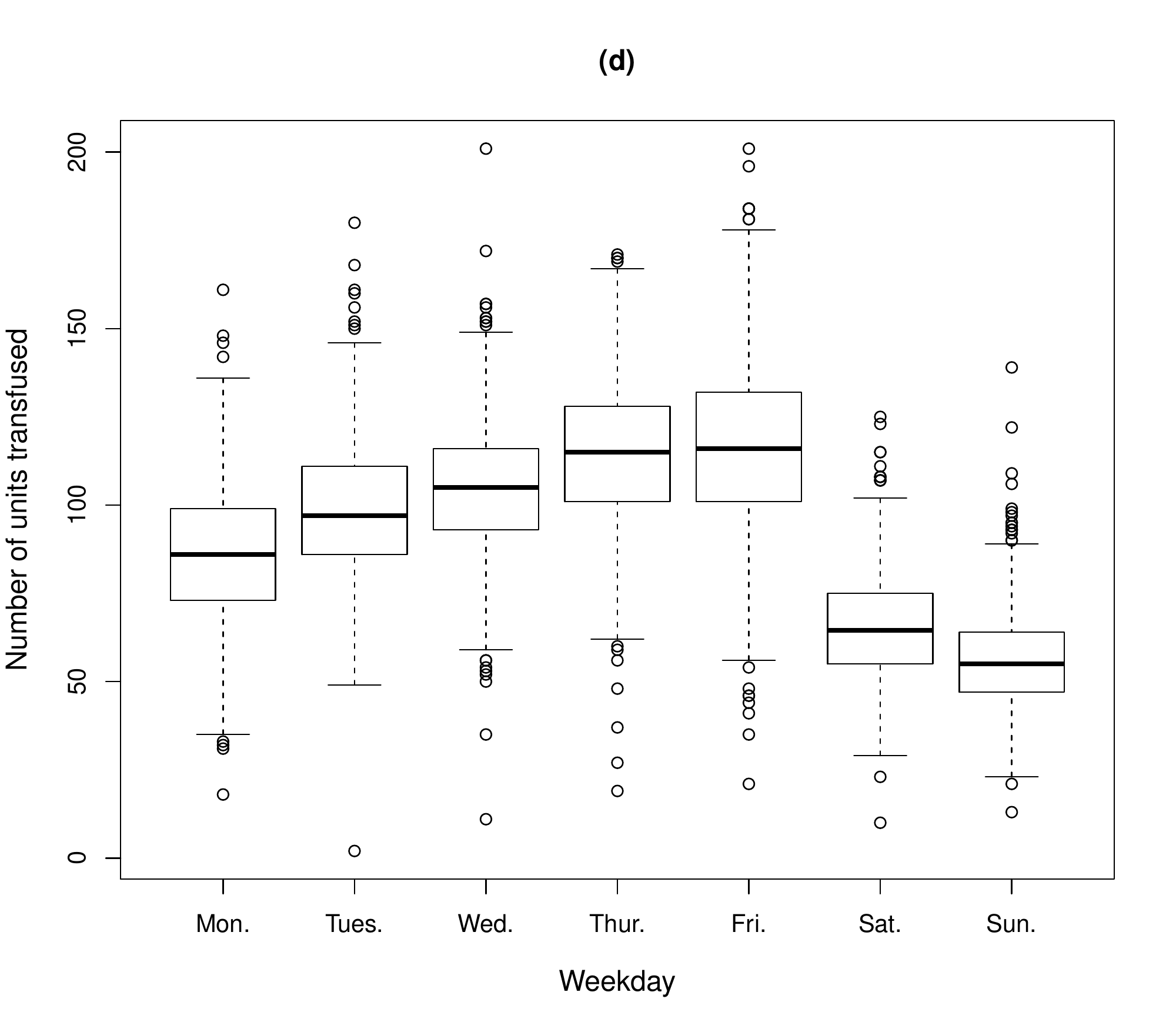}
}
\caption{Boxplots of difference (R-T) between units received and transfused (a) by year and (b) by day-of-week; boxplots of (c) number of units received by day-of-week and (d) number of units transfused by day-of-week}
\label{fig:RBCdemandvsordered}
\end{figure}

\noindent \textbf{Challenge 2: Large variation of the differences between ordered quantity and actual demand.} The mean and sd statistics for the daily number of units received, transfused and their difference are shown in Table \ref{demand_ordered}. As shown in the table, although the average difference between the number of units received (as a proxy of the order quantity by hospital blood banks on the previous day) and units transfused (actual demand) was close to zero, the standard deviation of the differences was very large. Figure \ref{fig:RBCdemandvsordered} (a) presents the boxplots of the differences between ordered quantity and actual demand by year from 2008 to 2018, which shows the ranges of the differences were wide for all the years and there was no observed trend across the years. In Figure \ref{fig:RBCdemandvsordered} (b), the boxplots of the differences reveals a strong day-of-week effect. The large variation of the differences was statistically significantly associated with the day-of-week effect and not significantly associated with the year (two-way ANOVA: among years $F=0.82, p=0.39$; across days of week, $F=204.86, p<0.001$). Figures \ref{fig:RBCdemandvsordered} (c) and (d) illustrates different patterns of number of units received and transfused by day-of-week. In Figure \ref{fig:RBCdemandvsordered} (c), the number of units received was much higher than the actual demand on Mondays and Fridays, and lower during weekends. This suggests hospital blood banks make larger orders on Sundays and Thursdays, whereas order levels on Fridays and Saturdays are lower due to low staffing levels. Figure \ref{fig:RBCdemandvsordered} (d) shows that the number of units transfused consistently increases from Monday to Friday, and then drops on weekends. Our demand forecasting algorithm takes into account the day-of-week effect and the clinical indicators that reflect the actual demand pattern leading to more accurate RBC demand prediction in support of ordering decisions.

\begin{table}[htbp]
\centering
\small
\begin{tabular}{|l|c|c|c|c|c|c|}
\hline
\textbf{Year} & \textbf{Number of units received (R)} & \textbf{Number of units transfused (T)} & \textbf{Difference (R-T)}  \\ \hline
2008 & 91.90 (63.20) & 89.51 (32.39) & 2.39 (46.96) \\ \hline
2009 & 88.65 (62.47) & 88.55 (30.02) & 0.10 (46.70) \\ \hline
2010 & 88.65 (62.05) & 88.66 (30.09) & -0.01 (46.52) \\ \hline
2011 & 96.79 (70.99) & 96.62 (31.05) & 0.17 (56.89) \\ \hline
2012 & 96.60 (69.38) & 95.80 (30.22) & 0.80 (54.42) \\ \hline
2013 & 93.25 (65.66) & 93.57 (29.74) & -0.32 (49.99) \\ \hline
2014 & 87.20 (59.56) & 86.96 (27.58) & 0.24 (45.97) \\ \hline
2015 & 86.75 (59.62) & 87.01 (27.29) & -0.26 (47.53) \\ \hline
2016 & 94.35 (63.86) & 94.05 (27.24) & 0.30 (51.42) \\ \hline
2017 & 97.39 (64.14) & 97.81 (27.83) & -0.41 (51.15) \\ \hline
2018 & 90.04 (61.43) & 93.05 (27.01) & -3.01 (49.10) \\ \hline
\end{tabular}
\caption{Mean (sd) of daily number of RBC units received (R), transfused (T), and their difference (R-T) by year in Hamilton hospital blood banks}
\label{demand_ordered}
\end{table}

\noindent \textbf{Challenge 3: Frequent same-day urgent orders.} We have described the process for routine ordering requests between hospital blood banks and CBS above. In urgent situations, non-scheduled orders that require same-day delivery can be placed. There are two types of urgent orders: ``as soon as possible" (ASAP) orders and ``STAT" orders. ASAP orders are usually dispatched by parcel express, and STAT orders are typically required in an emergency situation for bleeding patients, thus faster transportation methods such as taxis are used (the lead time for delivery of such orders is 2 to 3 hours to Hamilton, Ontario). From April 1st, 2012 to May 31st, 2015, of 1,156 calendar days (165 weeks), 1,012 (87.5\%) days had RBC deliveries from CBS to Hamilton blood banks\footnote{961 days excluding public holidays and Sundays during the period. That is, at least 51 deliveries happened on Sundays and/or holidays.}. Table \ref{ordertype} shows the number of RBC orders for each order type from Hamilton hospital blood banks to CBS during the period from April 1st, 2012 to May 31st, 2015. Among the days with deliveries, 322 (31.8\%) days, almost twice a week, had STAT orders, and 69 (6.8\%) days had ASAP orders. Among the 165 weeks, 90 (54.5\%) Fridays had same-day urgent orders (combined STAT and ASAP). From Table \ref{ordertype}, we have two observations: 1) There was a small number of days with same-day urgent orders but without routine orders. Of these orders, 37 (90.2\%) occurred on weekends; 2) The same-day urgent orders were mostly made on different days for different hospitals. Although the rate of same-day urgent orders may not be too concerning for individual blood banks, the pooled rate for all blood banks was significantly higher than the rate stratified by hospitals. This reflects the need for optimizing the inventory management as a network. Table \ref{ordercompare} presents the difference in means of demand patterns between dates with same-day urgent orders and dates with routine orders. The numbers of units received and transfused were much higher on dates with urgent orders, and there were significant differences of units transfused to trauma patients (doubled) and patients with abnormal laboratory tests (defined in Table \ref{variabledescription}).

\begin{table}[htbp]
\small
\begin{tabular}{|p{0.22\textwidth}|p{0.17\textwidth}|p{0.16\textwidth}|p{0.16\textwidth}|p{0.15\textwidth}|}
\hline
\textbf{Hospital blood bank*} & \textbf{Days with routine orders - n (\%)$^{\ddag}$} & \textbf{Days with STAT orders - n (\%)$^{\ddag}$} & \textbf{Days with ASAP orders - n (\%)$^{\ddag}$} & \textbf{Days with any order - n (\%)$^{\dag}$} \\ \hline
Hospital A    & 819 (96.6\%) & 102 (12.0\%) & 11 (1.3\%) & 848 (73.4\%)\\ \hline
Hospital B & 803 (95.9\%) & 104 (12.4\%) & 23 (2.8\%) & 837 (72.4\%) \\ \hline
Hospital C & 844 (96.8\%) & 72 (8.3\%) & 22 (2.5\%) & 872 (75.4\%)\\ \hline
Hospital D   & 789 (95.6\%) & 95 (11.5\%) & 21 (2.6\%) & 825 (71.4\%) \\ \hline
All hospitals   & 971 (95.9\%) & 322 (31.8\%) & 69 (6.8\%) & 1012 (87.5\%) \\ \hline
\end{tabular}%
\caption{Summary of RBC orders in Hamilton hospital blood banks from April 1st, 2012 to May 31st, 2015 \\
\small{* Hospitals A, B, C, and D represent the four hospital blood banks in Hamilton, Ontario. \\
$\ddag$ The denominators for the percentages of days with routine orders, STAT orders, and ASAP orders are the number of days with any order in the last column. \\
$\dag$ The denominator for the percentages of days with any order in the last column is 1,156 calendar days during the period.}}
\label{ordertype}
\end{table}

It is interesting to observe the issues of excess inventory levels and over-frequent same-day urgent orders simultaneously exist. The data shows hospital blood banks made same-day urgent orders when more patients were in severe condition even when there were units available in inventory, revealing a potential cognitive bias for overestimating shortage risks. These biases can be controlled using mathematical models for quantitative analysis. The integrated data-driven demand forecasting and inventory management model we propose can produce accurate demand predictions and generate robust ordering decisions based on historical data. It can significantly reduce the occurrence of ASAP and STAT orders while reducing inventory levels, resulting in significant savings with respect to both costs and resources.

\begin{table}[htbp]
\centering
\footnotesize
\begin{tabular}{|p{0.35\textwidth}|p{0.17\textwidth}|p{0.17\textwidth}|p{0.18\textwidth}|}
\hline
\textbf{} & \textbf{Dates with urgent orders - mean (sd)} & \textbf{Dates with routine orders - mean (sd)} & \textbf{Difference in means (95\% CI)} \\ \hline
Number of units received by blood bank & 119.35   (69.48) & 89.03   (67.09) & 30.32 (22.23 - 38.41) \\ \hline
Number of units transfused & 97.05   (29.71) & 86.71   (27.58) & 10.35 (6.94 - 13.75) \\ \hline
~~~~~Number of O units to Non-O recipients & 5.12   (5.03) & 3.99   (3.28) & 1.14 (0.62 - 1.66) \\ \hline
\multicolumn{4}{|l|}{Number of units transfused to} \\ \hline
~~~~~Patients with age between 18 and 75 & 61.09   (19.86) & 54.40   (18.17) & 6.69 (4.43 - 8.96) \\ \hline
~~~~~Trauma patients & 3.53   (8.42) & 1.74   (4.70) & 1.79 (0.94 - 2.64) \\ \hline
~~~~~Patients with abnormal creatinine & 29.23   (10.49) & 26.76   (10.04) & 2.46 (1.25 - 3.68) \\ \hline
~~~~~Patients with abnormal hemoglobin & 52.73   (14.43) & 49.07   (13.11) & 3.66 (2.01 - 5.30) \\ \hline
~~~~~Patients with abnormal INR & 88.64   (28.18) & 78.79   (26.14) & 9.85 (6.62 - 13.09) \\ \hline
~~~~~Patients with abnormal red cell width & 36.64   (13.94) & 33.22   (13.22) & 3.42 (1.81 - 5.03) \\ \hline
~~~~~Patients with abnormal pO2 & 15.52   (9.31) & 13.18   (7.97) & 2.33 (1.30 - 3.37) \\ \hline
~~~~~Patients at Juravinski Hospital & 39.25   (15.19) & 35.52   (14.18) & 3.73 (1.98 - 5.48) \\ \hline
~~~~~Patients at St. Joseph' Healthcare & 21.54   (10.80) & 18.87   (10.02) & 2.67 (1.43 - 3.91) \\ \hline
\end{tabular}
\caption{Comparison between dates with same-day urgent orders and dates with routine orders in Hamilton hospital blood banks}
\label{ordercompare}
\end{table}

\section{Literature Review}
\label{literature}

\subsection{Demand forecasting methods}
Most of the existing literature considers univariate demand forecasting for RBCs using time series or machine learning models. Salviano et al. \cite{Salviano2012, Salviano2013} developed an automatic application for demand forecasting, based on the Box and Jenkins (BJ) procedure. Their application forecasts the demands of blood components by using Seasonal AutoRegressive Integrated Moving Average (SARIMA) models to identify non-stationary and seasonal features. It can perform automatic order identification, parameter estimation and model validation to reduce human intervention and improve the efficiency of decision making for blood component distribution to hospitals. Kumari and Wijayanayake \cite{Kumari2016} proposed a model that manages the daily supply of platelets by forecasting the daily demand, where three forecasting techniques, moving average, weighted moving average, and exponential smoothing, were compared to minimize the shortage. Khaldi et al. \cite{Khaldi2017ArtificialNN} presented a case study of forecasting monthly demand of three blood components, RBC, plasma and platelets, using Artificial Neural Networks (ANNs). Guan et al. \cite{Guan} built a mathematical model to forecast short-term platelet usage and minimize wastage. Lestari and Anwar \cite{Lestari2017} investigated four methods to forecast blood demand involving moving average, weighted moving average, exponential smoothing, and exponential smoothing with trend using POM-QM software for supporting decision making of a blood transfusion unit in Indonesia. Among all the studies, Khaldi et al. \cite{Khaldi2017ArtificialNN} and Guan et al. \cite{Guan} appear to be the only two studies that have considered clinically-related indicators in short-term product-specific demand forecasting.

In this study, we train a hybrid demand forecasting model using a large number of clinical indicators, including patient characteristics, laboratory test results, patient diagnoses, and hospital locations. The model selects the most important clinical predictors for RBC demand forecasting to produce accurate forecasting results, generating valuable feedback of RBC utilization to CBS and hospitals.

\subsection{Operations research methods in blood demand and supply management}
Simulation models have been developed for BSCM, including blood collection, production, inventory and distribution. Mansur et al. \cite{Mansur} reviewed articles on BSCM from 1960 to 2017, and provided a concise summary of the articles based on four categories: blood product type, performance measurement, coordination hierarchy level, and blood inventory model. They point out that the solutions offered are not comprehensive and sometimes are difficult, if not impossible, to implement. They then used the blood management system in Indonesia as an example to suggest the need for a reliable inventory management system adaptive to demand fluctuation and blood supply pattern.

Beyond the articles surveyed in \cite{Mansur}, a number of additional references are pertinent for our proposed approach. Sirelson and Brodheim \cite{Sirelson1991} built a predictive model using simulation and linear regression that determines the outdate rate and the shortage rate as a function of the fixed base stock level and the mean daily demand, for blood banks with scheduled daily deliveries of platelet components from a regional blood centre. They showed that for blood banks with moderate to large mean demands there exist optimal base stock levels that can effectively keep the outdate rate and the shortage rate within favorable ranges. They also extended the model to distinguish the platelet demand by blood groups. Haijema et al. \cite{Haijema2007BloodPP} presented a combined Markov Decision Process and simulation approach with an application in a Dutch blood bank. Hemmelmayr et al. \cite{Hemmelmayr2010} established a two-stage stochastic optimization problem, which relies on sampling-based approaches involving integer programming to handle the stochastic demand and variable neighborhood search to improve computational efficiency. Zhou et al. \cite{Zhou2011} analyzed a periodic review inventory system for platelet components under two replenishment modes: regular orders placed at the beginning of a cycle, and expedited orders within the cycle characterized by an order-up-to level policy. They started with a single-item periodic review inventory system and then expanded their work to a multi-period inventory problem. They provided a numerical illustration and a sensitivity analysis using historical data, and showed that the optimal cost is significantly affected by demand uncertainty, lead times, seasonality, and age of expedited orders.

Although there have been many studies in the field of blood demand and supply management, the methods are typically developed based on various assumptions and are difficult to implement. Furthermore, to our knowledge, no study has considered integrating demand forecasting models into inventory management strategies for blood products. In this study, we investigate a multi-period inventory problem that mitigates the effects of forecasting errors from a data-driven demand forecasting model. The proposed integration strategy can help resolve practical challenges for RBC demand and supply chain management.

\subsection{Health science implementations}
There have been multiple inventory management approaches implemented in healthcare systems. Heitmiller et al. \cite{Heitmiller2010} was the first major study focusing on reducing blood wastage from the hospital side. They used the five-part Lean Sigma process, i.e., define, measure, analyze, improve, and control, to reduce RBC wastage with an emphasis on container wastage, where a control plan and a list of interventions by Lean Sigma were developed. They demonstrated there could be a 60\% reduction in RBC wastage with savings of more than \$800,000 over four years.

Kort et al. \cite{Kort2011} showed significant improvements, including a reduction of the median weekly outdating rate and a gain in the time until outdating, after implementing a combined approach of stochastic dynamic programming and simulation techniques. They stated the results brought confidence to personnel to apply and adopt the mathematical approach and the thrombocyte inventory management optimizer software tool. Collins et al. \cite{Collins2015} evaluated the effectiveness of multiple low-cost interventions that were implemented in January 2013 in the U.S., including educational outreach, print and digital messaging, and improved transportation and component identification modalities. They compared the RBC, platelet, and plasma wastage rates in the 16 months after these interventions with the rates prior to the interventions. They found significant decreases in the RBC and platelet wastage rates, however, there was an increase in the plasma wastage rate. The overall net cost savings of the reduced waste was estimated at \$131,520, excluding the intervention costs. Quinn et al. \cite{Quinn2019} designed and implemented a blood ordering algorithm, using a mathematical model based on the probability of RBC transfusion within 48 hours given certain hemoglobin levels, to provide a more accurate measure of RBC utilization. After implementation, they observed a significant reduction of the mean daily total RBC inventory level and the monthly RBC outdated units.

These applications showed significant cost savings could be achieved by applying mathematical modeling in BSCM. Our proposed integrated methodology framework, being data-driven produces more robust results, is generally applicable to a range of decision problems in BSCM, e.g., for different blood products. Moreover, to support accessibility and knowledge translation, we plan to develop a prototyping tool with a user-friendly interface using the proposed methods.

\section{Materials and Methods}
\label{method}
\subsection{Data description}
Our study dataset is constructed by processing the TRUST (``Transfusion Research for Utilization, Surveillance and Tracking") database from the McMaster Centre for Transfusion Research (MCTR). The TRUST database is a comprehensive database containing blood product, demographic, and clinical information on all hospitalizations at four Hamilton hospitals from April 2002 to the present. The database is updated monthly from two sources: the hospitals' Laboratory Information System (LIS) and Discharge Abstract Database (DAD). This study considers RBC inventory data and RBC transfusion-related clinical data in the TRUST database from 2008 to 2018. The study is approved by the Canadian Blood Services Research Ethics Board and Hamilton Integrated Research Ethics Board.

From 2008 to 2018, the study identifies 369,481 RBC transfusions for 60,141 patients in Hamilton. These consist of 236,856 transfusions to 39,811 inpatients and 132,625 transfusions to 21,581 outpatients. The patient characteristics (features) include age, gender, patient ABO Rh blood type, patient diagnosis, hospital facility, transfusion location, laboratory test, surgical procedure, and medication. We consider the following laboratory tests: hemoglobin (Hb), platelet count (PLTCT), creatinine, international normalised ratio (INR), red cell distribution width (RDW), immunoglobulins (IgG), mean platelet volume (MPV), mean corpuscular volume (MCV), white blood cell (WBC), mean corpuscular hemoglobin (MCHb), activated partial thromboplastin time (aPTT), fibrinogen, alanine aminotransferase (ALT), aspartate aminotransferase (AST), and blood gas (pO2, pCO2). RBC product-related features include mean transfusion age of blood, mean in-stock age, mean transfusion size, product ABO Rh type, expired rate, and destroyed rate. Hospital operations / policy related features include day of week, RBC transfusion compliance rate, and single unit issue rate.

The data for analysis is processed in two steps: 1) The dataset consists of all the transfused RBC units, and each row contains a unique RBC unit with product-related information and the transfusion recipient's characteristics as specified above. 2) A daily aggregated dataset is then constructed for demand forecasting. The dataset is organized by date, and each row contains the daily product and patient-related information. There are over 200 processed variables in the daily aggregated dataset using straightforward statistical transformations (e.g. mean, min, max, sum). Table \ref{variabledescription} presents a selected set of variables and their descriptions in the daily aggregated data. Variables are identified based on clinical relevance to RBC transfusions. Variables with over 70\% missing values are excluded, and methods of imputing missing values are based on the clinical definition and the use of the variable. For example, when a laboratory test value is missing, it may mean the test value is in the normal range so that a physician does not need to order additional laboratory tests. Outside of the laboratory test values, the percentage of missing values is low ($<$5\%) for the variables included. Visible errors in the database are corrected based on prior knowledge. For example, when the blood collection date is later than the expiry date or is greater than 42 days prior to the expiry date, the collection date is corrected to be the expiry date minus 42 days as the expiry date is considered accurate, coming from the CBS International Society of Blood Transfusion labelling (ISBT) label. Variables are normalized using the min-max method, i.e., $x_i=(z_i-\text{min}(z))/(\text{max}(z)-\text{min}(z))$.

All data preprocessing, analyses and modeling are performed using the R language for statistical computing \cite{R2018}. For the demand forecast model construction, data from 2008 to 2017 are designated for model training, and data in 2018 for test. Models are trained on the training dataset and cross-validation is used for hyperparameter tuning. Results are reported on the test dataset.

\begin{table}[htbp]
\footnotesize
\begin{tabular}{|p{0.15\textwidth}|p{0.75\textwidth}|p{0.1\textwidth}|}
\hline
\textbf{Name} & \textbf{Description}  & \textbf{Format} \\ \hline
ntransfusions & Number of RBC transfusions (units) & Integer \\ \hline
meanage    & Mean age in years of transfused patient  & Numeric  \\ \hline
female & Number of female patients transfused; sex is categorized by the biological attributes & Integer       \\ \hline
male & Number of male patients transfused; sex is categorized by the biological attributes & Integer       \\ \hline
patientABOgroup   & Number of transfused patients with  A / B / AB / O blood group, respectively & Integer     \\ \hline
patientRhtype     & Number of transfused patients with Positive / Negative  Rh type, respectively & Integer     \\ \hline
medicalgroup   & Number of transfused patients with medical diagnosis group: Medical intensive care unit (ICU) / Cardiac surgery / Non-cardiac surgery / Emergency room (ER) / Oncology / Trauma / Outpatient / Others, respectively  & Integer  \\ \hline
hospital       & Number of transfused patients in hospital: CMH (MUMC) / ML (Hamilton General) / HEND (Juravinski) / STJ (St. Joseph's) / WL (West Lincoln), respectively   & Integer  \\ \hline
location       & Number of transfused patients in location:  ICU / General medicine / Hematology / Cardiovascular surgery / General   surgery / Orthopedic surgery / Other surgery / Obstetrics \& Gynaecology / ER / Outpatient location, respectively & Integer \\ \hline
meanhb   & Mean hemoglobin before RBC transfusion; unit: g/L    & Numeric      \\ \hline
meanpltct   & Mean   platelet count before RBC transfusion; unit: x $10^9$/L  & Numeric  \\ \hline
meancreatinine     & Mean   creatinine before RBC transfusion; unit: $\mu$mol/L   & Numeric \\ \hline
meanINR  & Mean INR   before RBC transfusion & Numeric  \\ \hline
meanaptt & Mean   aPTT before RBC transfusion; unit: seconds      & Numeric   \\ \hline
abnormalhb  & Number   of patients with abnormal hemoglobin; abnormal is defined as \textless 80 g/L     & Integer \\ \hline
abnormalpltct & Number   of patients with abnormal platelet count; abnormal is defined as \textless 100 x$10^9$/L & Integer \\ \hline
abnormalcreatinine & Number   of patients with abnormal creatinine; abnormal is defined as \textgreater 90 $\mu$mol/L (female) and \textgreater 120   $\mu$mol/L (male)    & Integer \\ \hline
abnormalINR & Number   of patients with abnormal INR; abnormal is defined as \textless 1.6  & Integer   \\ \hline
abnormalRDW & Number   of patients with abnormal RDW; abnormal is defined as \textgreater 17\%  & Integer       \\ \hline
abnormalIgG & Number   of patients with abnormal IgG; abnormal is defined as \textless 7 g/L  & Integer  \\ \hline
abnormalMPV & Number   of patients with abnormal MPV; abnormal is defined as \textless 8 f/L  & Integer  \\ \hline
abnormalMCV & Number   of patients with abnormal MCV; abnormal is defined as \textless 80 f/L  & Integer \\ \hline
abnormalWBC & Number of   patients with abnormal WBC;   abnormal is defined as \textgreater 11 x$10^9$/L  & Integer \\ \hline
abnormalMCHb & Number   of patients with abnormal MCHb; abnormal is defined as \textless 29 pg & Integer   \\ \hline
abnormalaPTT & Number of   patients with abnormal aPTT; abnormal is defined as \textgreater 60 seconds & Integer \\ \hline
abnormalALT  & Number   of patients with abnormal ALT; abnormal is defined as \textless 17 U/L   & Integer \\ \hline
abnormalAST  & Number   of patients with abnormal AST; abnormal is defined as \textgreater 40 U/L  & Integer     \\ \hline
abnormalpO2  & Number   of patients with abnormal pO2; abnormal is defined as \textgreater 105 mmHg  & Integer   \\ \hline
abnormalpCO2 & Number   of patients with abnormal pCO2; abnormal is defined as \textgreater 45 mmHg & Integer \\ \hline
productABOgroup & Number of products with  A / B / AB / O blood group, respectively & Integer  \\ \hline
productRhtype &  Number of products with Positive / Negative  Rh type, respectively & Integer \\ \hline
meantranssize & Mean number of units transfused   per patient & Numeric \\ \hline
meanageofblood & Mean age in days of blood from   collection date to transfusion / expiry date & Numeric \\ \hline
meaninstockdays & Mean days in stock from received   by blood bank to transfusion / expiry date & Numeric \\ \hline
expiryrate & Rate of units expired; rate is calculated by number of units expired over total number of units per month. & Percentage \\ \hline
destroyedrate & Rate of units destroyed due to containment or other causes; rate is calculated by number of units destroyed over total number of units per month. & Percentage \\ \hline
weekday & Day of week: Monday / Tuesday /   Wednesday / Thursday / Friday / Saturday / Sunday & String \\ \hline
compliancerate & RBC transfusion compliance rate; RBC compliance is defined according to Choosing Wisely \cite{RBCguideline} and rate is calculated by number of transfusions with compliance over total number of transfusions per month. & Percentage\\ \hline
singleunitrate & Rate of transfusions with single unit issue; rate is calculated by number of transfusion episodes (single / multiple units issued at the same   time) with single unit issue over total number of transfusion episodes per day.  & Percentage\\ \hline
\end{tabular}
\caption{Data variable definition and description}
\label{variabledescription}
\end{table}

\subsection{Demand forecast modeling}
\label{demandforecastmethod}
\subsubsection{Model background}
\label{modelbackground}

In this study, we consider a combination of the Seasonal and Trend decomposition using Loess (STL) model and the eXtreme Gradient Boosting (XGBoost) model.

\noindent \textbf{STL model:} STL \cite{Cleveland1990} is a robust and efficient statistical method for time series decomposition. It can decompose a time series into three components, seasonality ($s_i$), trend ($\tau_i$), and residual ($e_i$) for time period $i$. The trend component is the long-term pattern that represents the increase or decrease in the time series over the observed period. The seasonality component represents a pattern of change that repeats itself over years at specific regular intervals less than a year, e.g., weekly, monthly, or quarterly. In this study, we only consider additive decomposition, thus the time series of RBC demand, denoted as $y_i$, can be written as $y_i=s_i+\tau_i+e_i$.

The main advantages of STL include its flexibility to handle different types of seasonality, the robust estimates of the trend and seasonal components (they are not affected by outliers), the ability to decompose time series with missing values, and fast computation. The rate of change of the seasonality and the smoothness of the trend-cycle can be controlled through two main parameters, the trend-cycle window (t.window) and the seasonal window (s.window). The parameter t.window is the span (in lags) of the Loess window for trend extraction and s.window is the span for seasonal extraction. Both should be odd numbers, and s.window should be at least $7$, according to Cleveland et al. \cite{Cleveland1990}. Smaller values yield greater sensitivity to detect changes. A value of s.window must be given, while, if omitted, a default value of t.window can be calculated using the number of periods and s.window. These parameters are very useful in this study since we observe irregular seasonality over time\footnote{This may be caused by blood production method changes, policy changes or operational changes.} for which the uncertainty could be better captured by STL than other time series models such as AutoRegressive Integrated Moving Average (ARIMA). On the other hand, STL has one critical drawback: STL is dependent on its own history. STL is not capable of capturing structural changes, such as non-linear patterns, that are associated with explanatory variables.

\noindent \textbf{XGBoost model:}  XGBoost \cite{xgboost} is a highly efficient and widely used machine learning model under the Gradient Boosting framework. In many data analytical challenges, it has proved to achieve state-of-the-art results \cite{xgboostgithub}. The idea of Gradient Boosting is to use smaller prediction models to build a more general model that fits the dataset used for training. In XGBoost, we use decision trees as the smaller prediction model. Decision trees are useful for separating a dataset into different categories, or leaves, according to decision criteria related to the dataset's distinct column values. Since we are working with a regression problem instead of classification, the leaves of the decision tree are not categories, but continuous-valued scores. The value predicted by the model is the sum of the scores pertaining to the leaves that were activated according to the predictor variables.

There are four unique features of XGBoost that make it a popular choice: 1) The weights in XGBoost are calculated by Newton's method, so it works fast as it does not require a line search. However, this also requires that the loss function of XGBoost must be twice continuously differentiable. 2) It can handle sparsity patterns, e.g. missing values, zero entries, in a unified way. 3) It has more regularization parameters to control the complexity of the model and prevent over-fitting. 4) It is highly flexible. It allows customized optimization objective functions and evaluation measures, as well as many hyperparameters for model tuning. Nonetheless, XGBoost is not good at dealing with time series data with long-term dependencies.

Thus, this leads us to develop a hybrid model combining the two methods. The combination of these two models can eliminate the drawbacks of each individual model while utilizing their advantages. It uses the STL model to extract the linear and seasonal components, and then allows the XGBoost model to handle the nonlinear patterns in the residuals. Similar hybrid models have been developed recently by combining ARIMA and machine learning models for various applications \cite{arimaxgboost2018, jiang2019}.

\subsubsection{Model development}

\begin{figure}[htbp]
\centering
\includegraphics[width=\textwidth]{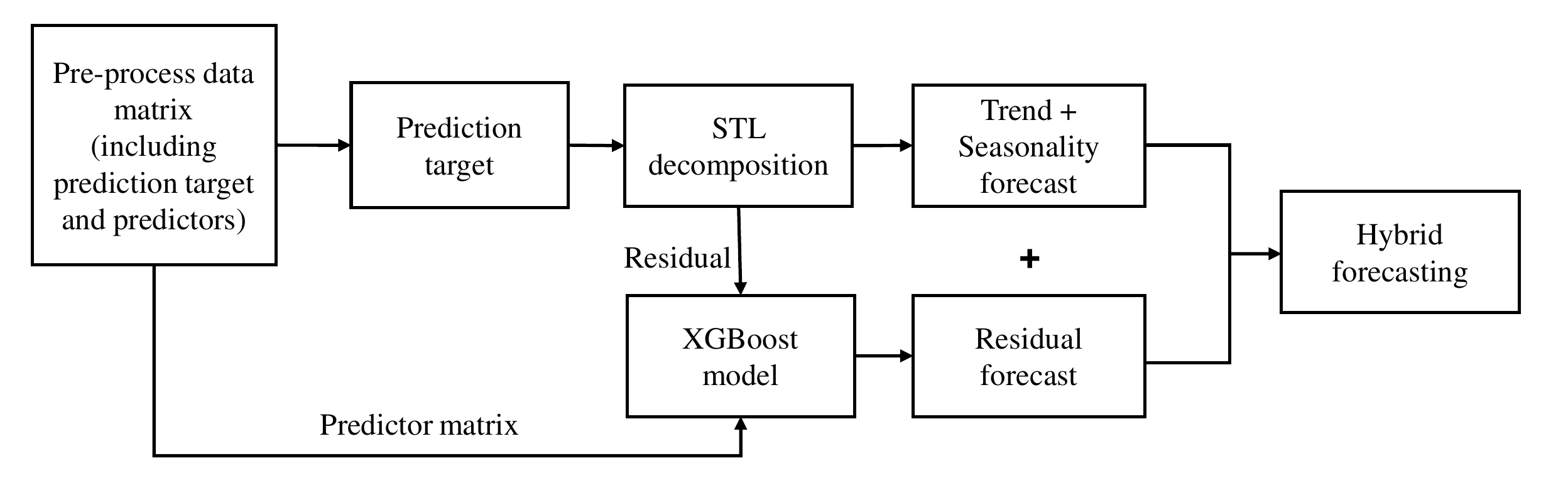}
\caption{STL + XGBoost hybrid forecasting algorithm for RBC demand}
\label{fig:hybridalgorithm}
\end{figure}

\noindent \textbf{Hybrid model: STL + XGBoost} Leveraging the advantages of the STL and XGBoost models, we propose an ensemble-based STL + XGBoost hybrid model for RBC short-term demand forecasting. As shown in Figure \ref{fig:hybridalgorithm}, the hybrid model starts with a time series decomposition of the daily RBC demand using an STL model, after which the STL residuals are forecast with an XGBoost model using a set of clinical predictors. The final forecast demand is the sum of the trend and seasonality components from the STL model and the predictions from the XGBoost model, written as
\begin{align}
\hat{y}_i &= s_i + \tau_i + \hat{e}_i \\
\text{where} ~~~~ \hat{e}_i&=\displaystyle\sum_{k=1}^Kf_k(x_i), ~f_k \in \mathcal{F}. \nonumber
\end{align}
The residuals forecast from the XGBoost model are denoted by $\hat{e}_i$, given the input predictors $x_i$. The space of regression trees is $\mathcal{F}=\{f(x)=w_{q(x)}\}$. The structure of each tree that maps the input predictors to the corresponding leaf index is represented by $q: \mathbb{R}^d \rightarrow \{1,\dots,\theta\}$, where $\theta$ is the number of leaves in the tree, and $w \in \mathbb{R}^{\theta}$ represents the weight of each leaf. Let $f_k$ correspond to an independent tree structure $q$ and leaf weights $w$. Let $\hat{e}_i^{(k)}$ be the prediction of the $i$-th instance at the $k$-th iteration. The objective function at the $k$-th iteration is
\begin{align}
& \mathit{L}^{(k)}=\displaystyle\sum_{i=1}^nl(e_i, \hat{e}_i^{(k-1)}+f_k(x_i))+\Omega(f_k) \\ \nonumber
& \text{where} ~~~~ \Omega(f_k)=\gamma\theta+\frac{1}{2}\lambda\|w\|^2.
\end{align}
The function $l$ is a differentiable convex loss function that measures the difference between the target $e_i$ and the prediction $\hat{e}_i^{(k-1)}$ at the ($k-1$)-th iteration added to the $f_k$ that most improves the model. A penalization term, denoted by $\Omega$, controls the complexity of the model to avoid over-fitting, where $\gamma$ is a penalty term on the number of leaves, so the larger $\gamma$ is the more conservative the model. Finally, $\lambda$ is a regulation term on the weights, increasing this value will make the model more conservative.

Define $g_i^{(k-1)}=\partial_{\hat{e}^{(k-1)}}l(e_i, \hat{e}_i^{(k-1)})$ and $h_i^{(k-1)}=\partial^2_{\hat{e}^{(k-1)}}l(e_i, \hat{e}_i^{(k-1)})$ as the first and second order gradient statistics on the loss function, and $I_j=\{i|q(x_i)=j\}$ as the instance set of leaf $j$. After applying the second order approximation and removing constant terms, the following loss function is minimized at step $k$,
\begin{align}
\label{lossfunction}
\hat{\mathit{L}}^{(k)}=\displaystyle\sum_{j=1}^\theta[(\displaystyle\sum_{i \in I_j}g_i^{(k-1)}) w_j+\frac{1}{2}(\displaystyle\sum_{i \in I_j}h_i^{(k-1)}+\lambda)w_j^2]+\gamma \theta.
\end{align}
For a fixed structure $q(x)$, the optimal weight $w^*_j$ of leaf $j$ is given by
\begin{align}
\label{weight}
w^*_j=-\frac{\sum_{i \in I_j}g_i^{(k-1)}}{\sum_{i \in I_j}h_i^{(k-1)}+\lambda}, ~~~~ j=1,\dots,\theta
\end{align}
and substituting (\ref{weight}) into (\ref{lossfunction}),
\begin{align}
\hat{\mathit{L}}^{(k)}(q)= -\frac{1}{2}\displaystyle \sum_{j=1}^{\theta}\frac{(\sum_{i \in I_j}g_i^{(k-1)})^2}{\sum_{i \in I_j}h_i^{(k-1)}+\lambda}+\gamma \theta.
\end{align}
This is the optimal loss for a fixed tree structure, however there might be thousands of possible trees. Instead of searching all possible tree structures, XGBoost uses a greedy algorithm to build a tree structure where the split is chosen with the maximum gain in the loss reduction. Let $I_L$ and $I_R$ be the instance sets for the left and right nodes, respectively. The gain in the loss reduction is calculated by
\begin{align}
\frac{1}{2}\left[\frac{(\sum_{i \in I_L}g_i^{(k-1)})^2}{\sum_{i \in I_L}h_i^{(k-1)}+\lambda}+\frac{(\sum_{i \in I_R}g_i^{(k-1)})^2}{\sum_{i \in I_R}h_i^{(k-1)}+\lambda}-\frac{(\sum_{i \in I_L \cup I_R}g_i^{(k-1)})^2}{\sum_{i \in I_L \cup I_R}h_i^{(k-1)}+\lambda}\right]-\gamma.
\end{align}

\subsubsection{Model training}
\label{modeltraining}
There are a number of hyperparameters for an XGBoost model, including learning rate, the maximum depth of a tree, the minimum sum of weights of all observations in a leaf, the fraction of observations to be randomly sampled for each tree, the fraction of columns to be randomly sampled for each tree, and the regularization term on weights. In addition, s.window and t.window for the STL model are also considered as hyperparameters in the hybrid model. The hyperparamters are tuned using grid search based on a pre-defined parameter space. The tuning process is measured by cross-validation on the training data and evaluated with the root mean squared error (RMSE). The optimal hyperparameters are selected with the minimum RMSE using 5-fold cross-validation.

Variable selection proceeds in an iterative manner based on the variable importance scores calculated from the XGBoost models. Variable importance is the relative improvement in the performance measure contributed by a variable weighted by the number of observations for each decision tree, then averaged across all the decision trees within the model \cite{bookHastie2009}. We initialize the iterative process with a model using all variables on the training dataset and evaluate the model performance measure on the test dataset, then select a subset of variables based on a pre-defined variable importance threshold for the next iteration. We repeat the process until there is no improvement observed in the performance measure, then the subset with the most important variables is finalized. The variables reported in the result section are the final set of variables selected through this process.

The prediction target for the demand forecasting model is the daily RBC demand. In order to reduce costs, a second target, semiweekly demand, is considered. The semiweekly demand is defined as the three-day demand for Tuesday, Wednesday and Thursday, and the four-day demand for Friday, Saturday, Sunday and Monday. The semiweekly demand is calculated from the daily demand prediction.

\subsubsection{Model evaluation}
We evaluate the model performance using the following two accuracy measures: Mean Absolute Percentage Error (MAPE) and Root Mean Square Error (RMSE). We report the values of MAPE and RMSE of the trained hybrid model on the test data. The model performance of the hybrid model is compared to a single STL model, an STL + linear regression model, and a long short-term memory (LSTM) model. An LSTM model \cite{Sak2014LongSM} is one type of recurrent neural network model used in the field of deep learning. It is well suited to handle long-term dependencies in time series data and has many applications, such as time series anomaly detection, short-term traffic forecast, and handwriting recognition.

\subsection{Integrated ordering strategy for RBC inventory management}
\label{section:ordering}
We consider a multi-period inventory problem for RBC units with fixed shelf life in a rolling horizon framework. The model assumptions are as follows: i) Based on our data, we assume a fixed duration of 10 days from blood collection date to received date resulting in a shelf life of 32 days for units arriving at hospital blood banks. Usually, the first two days after collection are spent on testing in blood production sites at CBS and the units are available for distribution on the third day, however, this could vary due to many reasons. ii) The RBC issuing policy in hospital blood banks is assumed to be a First-In First-Out (FIFO) withdrawal policy. iii) We assume an infinite supply of RBC units at CBS, so that all orders made from hospital blood banks to CBS can be fulfilled \footnote{This assumption is verified by CBS. CBS confirmed that over 98\% of hospital blood bank orders were satisfied based on their data.}.

The order of events occurring in each period is given as follows: i) Based on the inventory policy used, an order is placed (if necessary) at the end of each period. A routine delivery cost is charged for every order. ii) Fresh units arrive at the beginning of each period according to the quantity ordered at the end of the previous period. The inventory level of each age is then updated. iii) The demand is observed and satisfied as much as possible. If there is a shortage, an urgent delivery order for the unmet demand is requested and the unmet demand is satisfied during the same period. iv) At the end of each period, the remaining inventory is carried over to the next period. Expired units are discarded. Wastage costs are charged for expired units and, if applicable, same-day urgent delivery costs are charged.

Usually, an optimization problem to determine the ordering strategy is set up with prior assumptions on the demand and supply distributions \cite{Birge2011}. Recently, Bertsimas and Kallus \cite{Bertsimas2020} considered a conditional stochastic optimization problem given imperfect observations, and developed a framework to prescribe optimal decisions using observed explanatory variables. In this study, since our goal is not only to propose an ordering decision but also to develop an accurate model that identifies clinical predictors for RBC demands, we do not formulate the optimization problem using observed clinical indicators directly, as in Bertsimas and Kallus \cite{Bertsimas2020}. Instead, we proceed with the demand estimation and ordering strategy optimization in two separate steps. The structure of this data-driven inventory management problem is consistent with the structure for a newsvendor policy proposed in Huber et al. \cite{Huber2019}.

As described in Section \ref{demandforecastmethod}, we have developed a demand forecasting model to predict future RBC demand using clinical predictors. This provides important information to hospital blood banks and blood suppliers for model generalisation and knowledge translation. Multiple techniques have been applied to improve model accuracy, however, the existence of forecasting errors cannot be avoided. In the inventory optimization step for ordering decisions, we propose an ordering strategy considering two extra decision variables to control the cumulative loss due to the forecasting errors from the demand predictions of the hybrid model for this mutli-period inventory problem: inventory target ($S$) and reorder level ($s$). The proposed ordering strategy is a modified version of classical $(s, S)$ policies \cite{Scarf1959THEOO}. The inventory target, $S$, defines an upper limit on the inventory level to avoid excessive inventory levels that may arise from demand over-estimation. The reorder level, $s$, sets a lower limit on the inventory level to avoid the need for urgent deliveries that may arise from demand under-estimation. Under this policy, the order quantity is driven by the demand prediction from the hybrid model controlled by the following criteria: When the inventory level is below the reorder level, the order quantity is at least the number of units to bring the inventory level back to the reorder level, but cannot make the inventory level greater than the inventory target; if the inventory level is above the reorder level, no order is required. Thus, the final ordering strategy is a function of the predictions from the hybrid demand forecasting model, the inventory target, and the reorder level.

The age of an RBC unit is denoted by $m \in \{1,\dots,M\}$. Let $a, h, u, w$ be the routine delivery cost (per order), unit inventory holding cost, unit urgent delivery cost, and unit wastage cost for each expired unit, respectively. Let $I_{i}^{m}$ denote the inventory level of units of age $m$ at the end of period $i \in {1,\dots,T}$. The remaining demand after withdrawing all products having ages from $m$ to $M$ using the FIFO withdrawal policy is denoted by $R_i^{m}$. The order quantity is denoted as $z_i$ in each period $i \in {1,\dots,T}$, and $\mathbb{I}(z_i>0)$, is the indicator that an order occurs in period $i$. At the end of period $i$, the actual demand $y_i$ is updated. Then, the inventory level $I_{i}^{m}$ for each age $m$, units required for urgent delivery $B_i$, and wasted items due to expiration $I_{i}^{M}$ are calculated. The proposed cost function is given as follows:
\begin{align}
\label{obj1}
c(z_i) &= a~\mathbb{I}(z_i>0)+h(\sum_{m=1}^{M-1}I_i^m)+uB_i+wI_i^M, ~~~ \text{for}~~ i=1,\dots,T,
\end{align}
where
\begin{align}
\label{con3}
R_i^m &= (y_i-\sum_{j=m}^{M-1}I_{i-1}^j)^{+}, ~~~~~ m=1,\dots,M \\
\label{con4}
I_{i}^m &= (I_{i-1}^{m-1}-R_i^{m})^+,  ~~~~~ m=2,\dots,M \\
\label{con5}
I_{i}^1 &= (z_{i}-R_i^1)^+, \\
\label{con6}
B_i &= (\sum_{j=1}^{M-1}I_{i-1}^j - \sum_{j=1}^MI_i^j+z_{i}-y_i)^{+}.
\end{align}
The cost function in equation (\ref{obj1}) includes four types of costs: routine delivery, inventory holding, urgent delivery and wastage cost over the planning horizon of $T$ periods. Equation (\ref{con3}) defines $R_i^{m}$ according to the FIFO withdrawal policy. Equations (\ref{con4}) and (\ref{con5}) define the inventory dynamics. Equation (\ref{con6}) calculates the number of units requiring urgent delivery. In the model, all variables are non-negative. The average cost can be written as $E[c(z)]=\frac{1}{T}\sum_{i=1}^{T} c(z_i)$.

We propose an integrated ordering strategy where the order quantity is the predicted demand, $\hat{y}_i$, from the hybrid model in Section \ref{demandforecastmethod} controlled by an optimal inventory target, $S^*$, and an optimal reorder level, $s^*$. The optimal inventory target and reorder level values are learned through the training data by minimizing the difference between the average costs under the predicted demands and the actual demands as a prior \cite{Shi2016}, rather than using the classical ordering structure based on estimated demand distribution \cite{Scarf1959THEOO, Ban2020}. We set the initial inventory, $I_0$, to be the mean inventory level according to the first three months of data. We assume this is an inventory level that the decision makers at hospital blood banks can accept, but it can be adjusted if needed  (in particular the effect of lowering the initial inventory could be explored). Let $I_i$ denote the aggregate inventory level of all non-expired units for period $i$, such that, $I_i=\sum_{j=1}^{M-1}I_{i}^j$ for $i=1,\dots,T$. Let $S^*$ be the optimal inventory target and $s^*$ be the optimal reorder level. The procedure to generate the ordering strategy is described as follows:
\begin{enumerate}
\item \textbf{Determine the optimal inventory target, $S^*$}: First, we calculate the cost for each period using the actual demand, $y_i$, as the ordering decision, $z_i$, and denote the average cost as $E[c(y)]=\frac{1}{T}\sum_{i=1}^{T} c(y_i)$ over the planning horizon of $T$ periods. When the order quantity is equal to the actual demand, the inventory level is always the same as the initial inventory level for all time periods. In reality, it is not possible to know the actual demand in advance. Ordering according to the actual demand at a given initial inventory level is used as a gold standard for obtaining the optimal inventory target. Second, we calculate the cost of each period using the predicted demand bounded by the inventory target, defined as $\min(\hat{y}_i, S-I_{i-1})$, as the ordering decision. We denote the average cost as $E[c(\hat{y}, S)]=\frac{1}{T}\sum_{i=1}^{T} c(\hat{y}_i, S)$ under different $S \in \Xi$ values, where $\Xi$ is the feasible set of $S$. The optimal inventory target, $S^*$, is determined by minimizing the absolute difference between the two average costs, such that, $S^*=\arg \min_{S \in \Xi} \left\|E[c(y)]-E[c(\hat{y}, S)]\right\|$.
\item \textbf{Seek the optimal reorder level, $s^*$:} Using the optimal inventory target as an input variable, we consider the decision variable as the predicted demand bounded by the optimal inventory target, $S^*$, and the reorder level, $s \in \xi$, where $\xi$ is the feasible set of $s$. If $s>I_{i-1}$, the order quantity is the maximum of $\min(\hat{y}_i, S^*-I_{i-1})$ and $(s-I_{i-1})$; else the order quantity is zero. The average cost is denoted as $E[c(\hat{y}, S^*, s)]$. The optimal reorder level, $s^*$, is determined by minimizing the absolute difference between $E[c(y)]$ and $E[c(\hat{y}, S^*, s)]$, such that, $s^*=\arg \min_{s \in \xi} \left\|E[c(y)]-E[c(\hat{y}, S^*, s)]\right\|$.
\item \textbf{Generate the proposed order quantity, $z_i^*$}: The proposed ordering decision integrating the prediction generated by the hybrid demand forecasting model, the optimal inventory target, and the optimal reorder level can be calculated as follows: For $i=1,\dots,T$,
\begin{align}
\label{orderingpolicy}
 & \text{if} ~~ I_{i-1} < s^*, \quad z_{i}^*={\begin{cases}
        S^*-I_{i-1} \quad\quad \text{if} ~~ \hat{y}_i > S^*-I_{i-1},\\
        \hat{y}_i \quad\quad\quad\quad~~~  \text{if} ~~ s^*-I_{i-1} \le \hat{y}_i \le S^*-I_{i-1},\\
        s^*-I_{i-1} \quad\quad \text{if} ~~ \hat{y}_i<s^*-I_{i-1},\\
     \end{cases}} \\
 & \text{else}  \quad z_{i}^*=0.  \nonumber
\end{align}
\end{enumerate}

The procedure is applicable to both the daily and the semiweekly demand predictions. The optimal inventory target is determined by errors in the daily demand predictions. Under a given optimal inventory target, since the reorder level is used to control the inventory variations due to forecasting errors, the calculation of the optimal reorder levels for the daily and semiweekly predictions are performed separately.

\section{Results}
\label{result}

\subsection{Demand forecasting}
\label{forecastresult}

\begin{figure}[htbp]
\centering \makebox[\textwidth]{
\includegraphics[width=\textwidth]{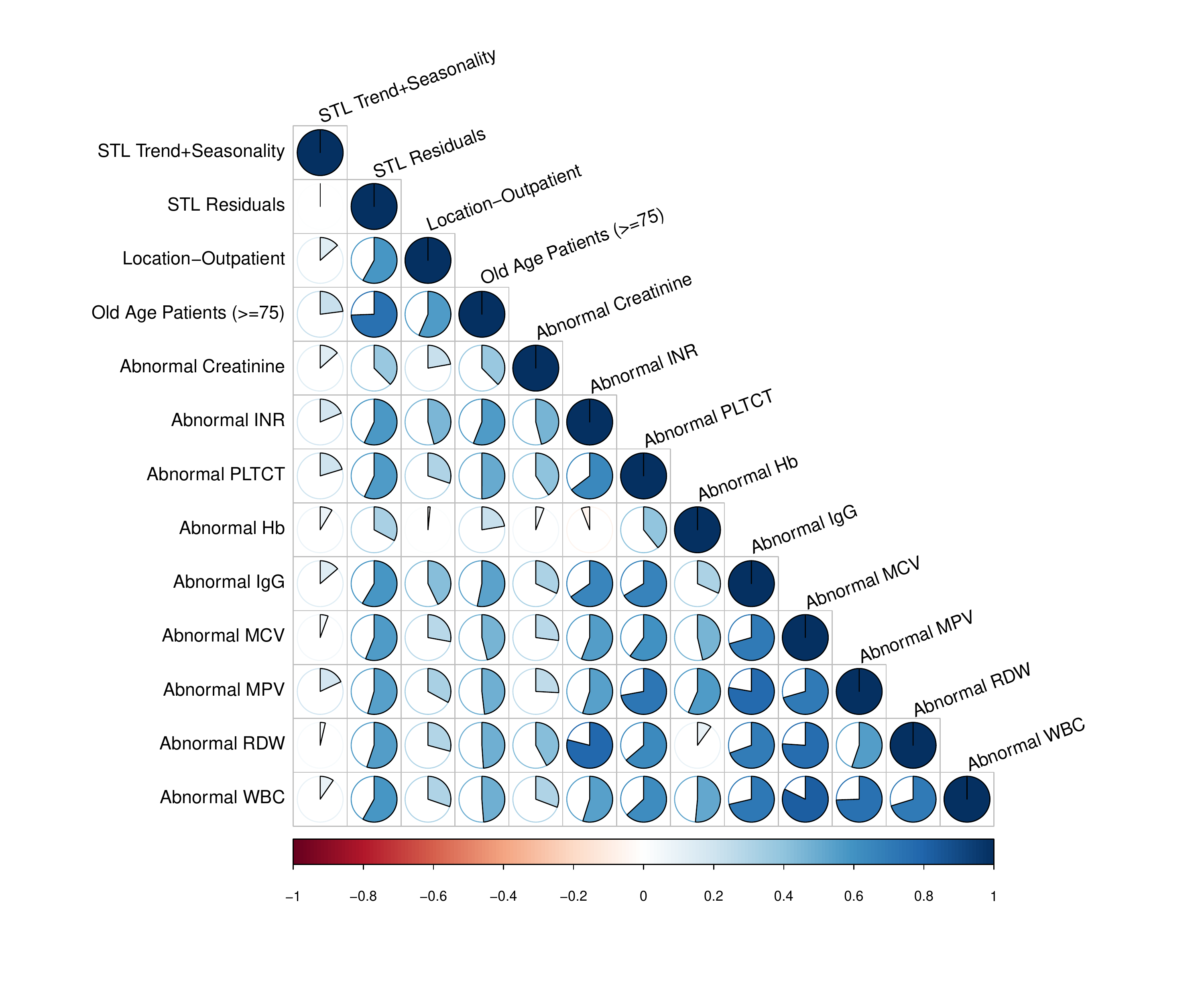}
}
\caption{Spearman's rank correlation among STL Trend + Seasonality, STL residuals, and selected clinical predictors (The pie shape in each cell represents the correlation statistics - the larger the shaded area of the pie shape, the higher the estimated correlation. The shape color is coded from the highest negative correlation -1 [dark red] to the highest positive correlation +1 [dark blue] as shown in the heat map legend at the bottom. In this figure, all of the shapes are blue representing positive correlations.)}
\label{fig:correlation}
\end{figure}

A statistical summary of selected variables is presented in Table \ref{statisticalsummary}. As described previously, the demand forecasting model is trained on the training dataset, and evaluated on the test dataset. The mean (sd) of the daily number of transfused units in the training dataset was 91.85 (29.62), while the mean (sd) in the test dataset was 93.05 (27.01). Over the entire time period, the estimated Sen's slope of daily RBC demand was 0.017 (95\% confidence interval [CI]: -0.036, 0.075) units per month. Although no statistically significant trend was observed over the entire period (Mann-Kendall trend test: $p=0.480$), change points in the univariate time series were detected using the non-parametric E-Divisive with Medians (EDM) algorithm \cite{James2016}. The mean (CI) of daily transfused units before 2014-01-01, during 2014 and 2015, and after 2016-01-01 were 92.08 (90.80, 93.37), 86.99 (84.99, 88.98), and 95.01 (93.39, 96.63), respectively. This reflects that the trend of daily transfused units is nonlinear over time. A very small number of outliers, 0.25\% of days, were detected, where the outliers were identified as the data points lying outside of 1.5 times of the interquartile range (IQR). Different seasonality patterns, day of week or month effects, were observed when stratified by year using multilevel mixed-effects models. These observations confirm the suitability of an STL model to deal with the changes in nonlinear trend and seasonality patterns. After time series decomposition using STL, we find significantly higher correlations between the STL residuals and selected clinical predictors than the correlations between the STL Trend + Seasonality and those predictors, as shown in Figure \ref{fig:correlation}. Moreover, Figure \ref{fig:correlation} shows the correlations were high among the abnormal laboratory test results. As described in Section \ref{modelbackground}, XGBoost models are well equipped to capture nonlinear structures and correlations in the multivariate analysis. Hence, the construction of a hybrid STL + XGBoost model is justified.

\begin{table}[htbp]
\small
\begin{tabular}{|l|c|c|}
 \hline
\textbf{Variable} & \textbf{~~~Daily mean (sd)~~~} & \textbf{~~~~~~Daily range~~~~~~} \\ \hline
Number of units transfused & 92.43 (28.27) & (18, 201) \\ \hline
Number of units received by blood bank & 103.71 (69.49) & (0, 321) \\ \hline
\multicolumn{3}{|l|}{Number of patients (transfused and non-transfused) with   abnormal} \\ \hline
~~~~~MPV & 755.44   (221.75) & (130,   1236) \\ \hline
~~~~~RDW & 297.42   (80.21) & (33,   475) \\ \hline
~~~~~IgG & 12.78   (9.79) & (0,   43) \\ \hline
~~~~~INR & 518.30   (126.06) & (154,   830) \\ \hline
~~~~~Creatinine & 104.31   (21) & (47,   175) \\ \hline
~~~~~Hemoglobin (Hb) & 157.31   (46.5) & (48,   322) \\ \hline
~~~~~Platelet count (PLTCT) & 209.39   (43.6) & (71,   337) \\ \hline
~~~~~MCV & 84.37   (27.75) & (12,   163) \\ \hline
~~~~~WBC & 404.27   (66.46) & (74,   568) \\ \hline
\multicolumn{3}{|l|}{Number of units transfused to patients with abnormal pre-transfusion} \\ \hline
~~~~~MPV & 26.9 (25.36) & (0, 116) \\ \hline
~~~~~RDW & 34.67 (13.28) & (4, 91) \\ \hline
~~~~~IgG & 1.03 (1.56) & (0, 16) \\ \hline
~~~~~INR & 85.1 (26.89) & (15, 180) \\ \hline
~~~~~Creatinine & 27.51 (10.16) & (2, 78) \\ \hline
~~~~~Hemoglobin (Hb) & 54.45 (15.3) & (13, 120) \\ \hline
~~~~~Platelet count (PLTCT) & 28.56 (10.42) & (4, 94) \\ \hline
~~~~~MCV & 8.12 (4.95) & (0, 34) \\ \hline
~~~~~WBC & 28.52 (10.07) & (4, 74) \\ \hline
\multicolumn{3}{|l|}{Number of units transfused at} \\ \hline
~~~~~General   medicine location & 31.7 (13.62) & (2, 116) \\ \hline
~~~~~Intensive care   unit & 12.31 (9.11) & (0, 105) \\ \hline
~~~~~Hematology   location & 12.53 (5.32) & (0, 47) \\ \hline
~~~~~Cardiovascular   surgery location & 8.16 (6.06) & (0, 40) \\ \hline
~~~~~General   surgery location & 9.07 (6.96) & (0, 73) \\ \hline
~~~~~Orthopedic   surgery location & 3.57 (3.25) & (0, 35) \\ \hline
~~~~~Outpatient   location & 4.85 (6.2) & (0, 32) \\ \hline
~~~~~Emergency room & 4.26 (3.58) & (0, 26) \\ \hline
~~~~~Other surgery   location & 4.78 (4.21) & (0, 42) \\ \hline
~~~~~Obstetrics and   gynecology location & 1.19 (2.17) & (0, 30) \\ \hline
~~~~~Trauma & 2.21 (6.7) & (0, 96) \\ \hline
\multicolumn{3}{|l|}{Number of units transfused to patients with} \\ \hline
~~~~~Inpatient & 62.85 (17.64) & (14, 145) \\ \hline
~~~~~Age \textgreater{}=75 & 29.08 (12.43) & (2, 98) \\ \hline
~~~~~Age between 18   and 74 & 59.04 (19.64) & (13, 156) \\ \hline
~~~~~Female & 41.3 (15.33) & (4, 105) \\ \hline
~~~~~Male & 51.14 (17.87) & (5, 129)  \\ \hline
\end{tabular}
\caption{Statistical summary of selected variables}
\label{statisticalsummary}
\end{table}

\begin{figure}[htbp]
\centering \makebox[\textwidth]{
\includegraphics[width=\textwidth]{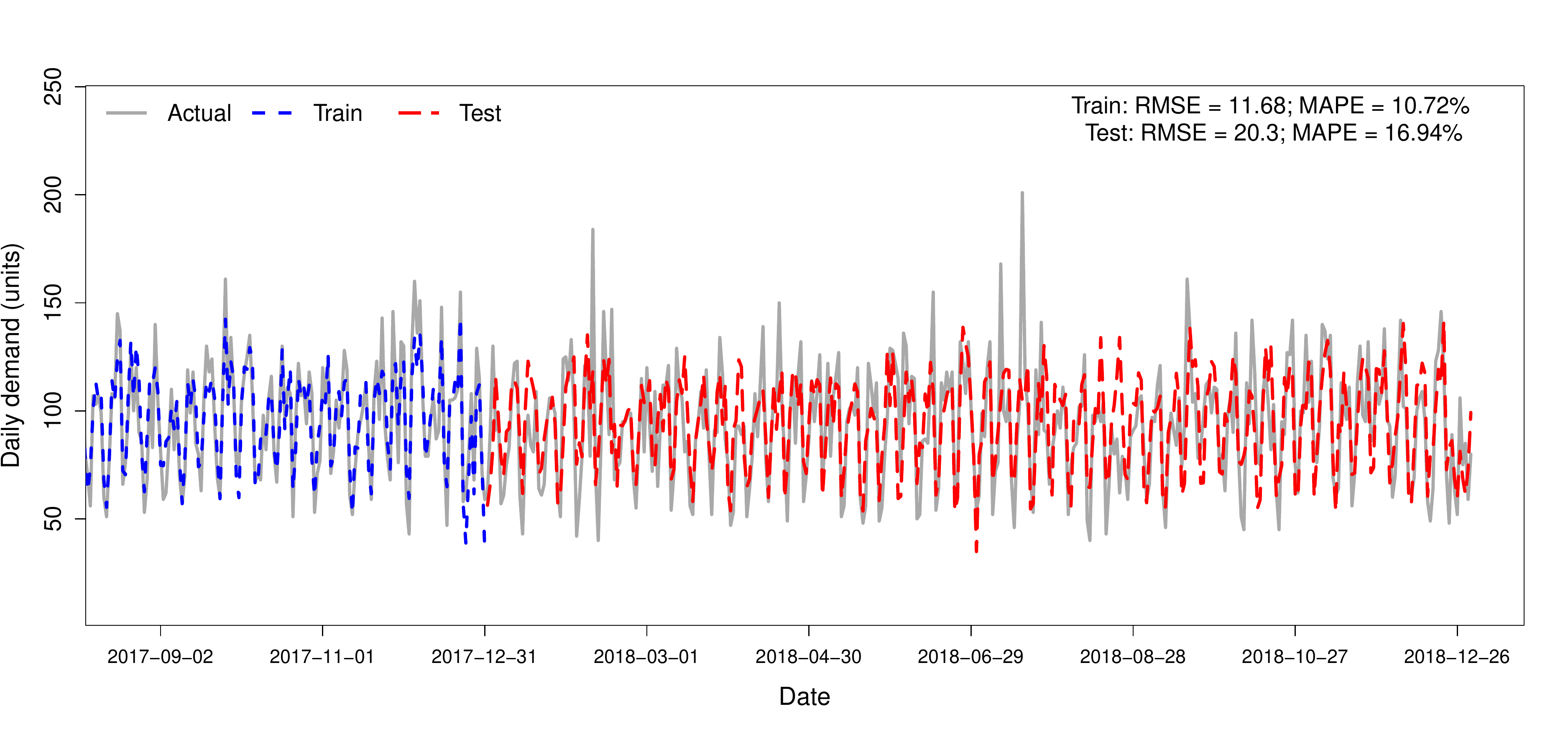}
}
\caption{Predicted RBC demand versus actual demand}
\label{fig:predictdemand}
\end{figure}

Figure \ref{fig:predictdemand} shows model performance of daily RBC demand predictions using the proposed hybrid model. The grey line represents the actual daily demand from the data, the blue dashed line shows the model predictions on the training dataset (partial), and the red dashed line shows the demand forecasts for the trained model on the test dataset. The variable selection and hyperparameter tuning processes are performed in an iterative manner. The mean (sd) of the predicted daily demand on the test dataset is 94.85 (21.72) units, and the mean (sd) of the true daily demand is 93.05 (27.01) units. The final model selected has an RMSE of 20.3 and a MAPE of 16.94\% on the test dataset. Table \ref{tab:comparemodels} presents the performance of the proposed hybrid model, a single STL model, an STL + linear regression model, and an LSTM model. The accuracy of the hybrid model is significantly higher than the single STL model, indicating that the inclusion of the multivariate analysis using clinical predictors produces more accurate RBC demand forecasts. When replacing the XGBoost model with a linear regression model, the accuracy decreases, reflecting that XGBoost can better handle the nonlinear patterns in the data. There is no significant improvement between the performance of the hybrid model and the LSTM model. However, compared to a tree-based model, such as XGBoost, building and tuning an LSTM model requires in-depth knowledge of neural network models. The proposed STL + XGBoost model has a simpler model structure which is much simpler to apply, and is better able to produce statistical inference.

\begin{table}[htbp]
\small
\begin{tabular}{lllll}
\hline
\multicolumn{1}{|l|}{\multirow{2}{*}{}} & \multicolumn{2}{c|}{Train} & \multicolumn{2}{c|}{Test} \\ \cline{2-5}
\multicolumn{1}{|l|}{} & \multicolumn{1}{l|}{RMSE~~~} & \multicolumn{1}{l|}{MAPE~~~} & \multicolumn{1}{l|}{RMSE~~~} & \multicolumn{1}{l|}{MAPE~~~} \\ \hline
\multicolumn{1}{|l|}{STL + XGBoost} & \multicolumn{1}{l|}{11.68} & \multicolumn{1}{l|}{10.72\%} & \multicolumn{1}{l|}{20.3} & \multicolumn{1}{l|}{16.94\%} \\ \hline
\multicolumn{1}{|l|}{STL Alone} & \multicolumn{1}{l|}{26.51} & \multicolumn{1}{l|}{23.22\%} & \multicolumn{1}{l|}{30.28} & \multicolumn{1}{l|}{25.06\%} \\ \hline
\multicolumn{1}{|l|}{STL + Linear regression} & \multicolumn{1}{l|}{18.73} & \multicolumn{1}{l|}{17.22\%} & \multicolumn{1}{l|}{21.67} & \multicolumn{1}{l|}{18.56\%} \\ \hline
\multicolumn{1}{|l|}{LSTM} & \multicolumn{1}{l|}{16.38} & \multicolumn{1}{l|}{12.58\%} & \multicolumn{1}{l|}{21.18} & \multicolumn{1}{l|}{16.50\%} \\ \hline
\end{tabular}
\caption{Model performance comparison}
\label{tab:comparemodels}
\end{table}

\begin{figure}[htbp]
\centering \makebox[\textwidth]{
\includegraphics[width=\textwidth]{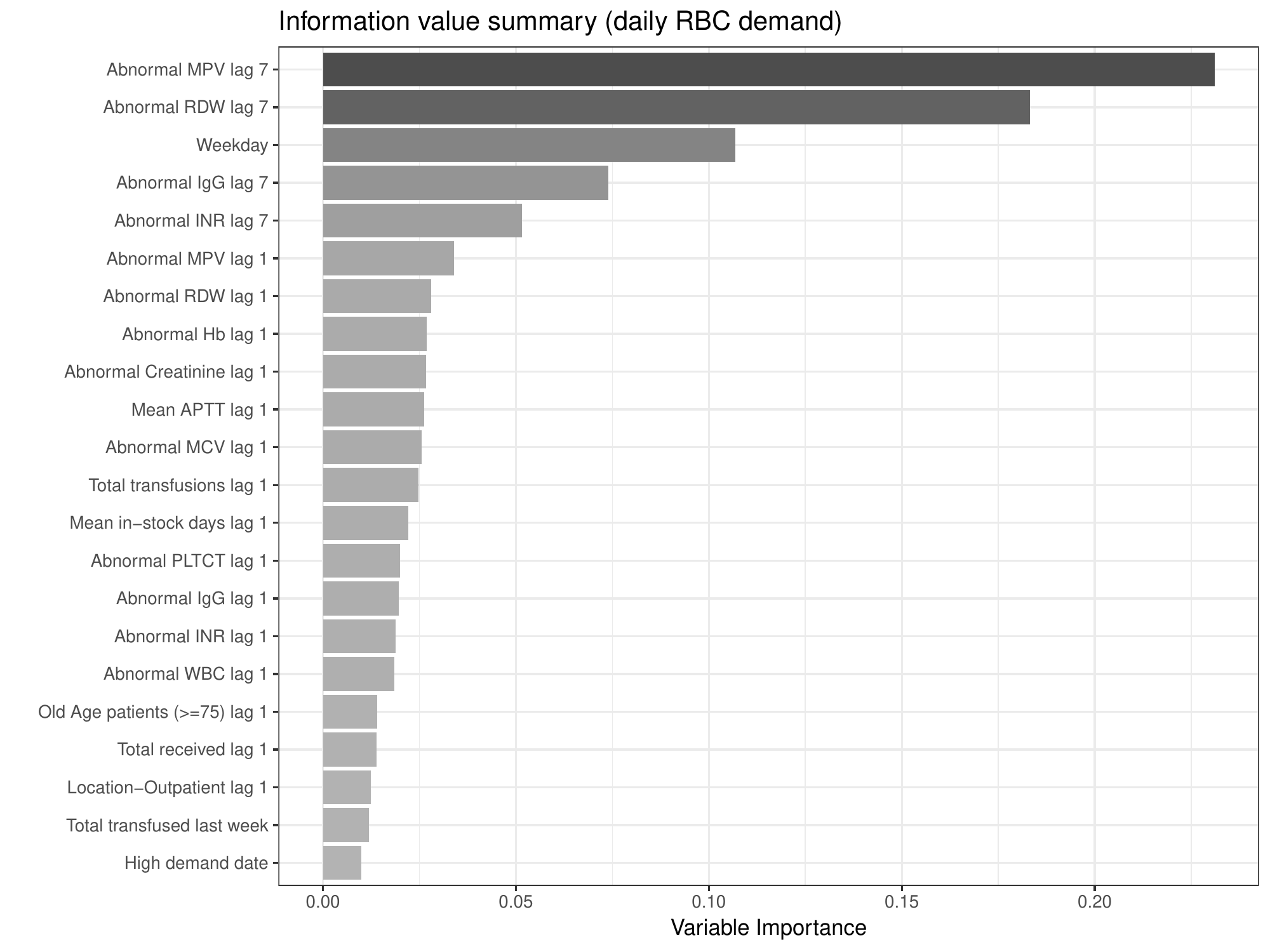}
}
\caption{Variable importance for daily RBC demand}
\label{fig:variableimportance}
\end{figure}

Figure \ref{fig:variableimportance} presents the relative variable importance of all 22 daily RBC demand predictors, selected after the iterative variable selection process described in Section \ref{modeltraining}, from highest to lowest. Among the top five predictors, it is not surprising that the importance of the weekday variable is high since we have observed significant day of week effects. Interestingly, four variables refer to the number of patients with abnormal laboratory test results that occurred seven days ago (lag 7). The next 15 variables capture the patient characteristics and RBC inventory on the previous day (lag 1). The second-to-last variable reflects the total RBC demand of the previous week, and the last variable refers to the dates with high daily demands, helping to correct the minor outlier effects.

As shown in Figure \ref{fig:correlation}, there exist significant positive correlations between the STL residuals and the laboratory test results, respectively. A laboratory test result can be used for condition evaluation within seven days from the date of specimen collection. The data shows that among the transfused patients, 55.72\% had abnormal MPV values at lag 1 and 55.18\% at lag 7; 39.08\% had abnormal RDW values at lag 1 and 51.64\% at lag 7; 59.95\% had abnormal IgG values at lag 1 and 58.62\% at lag 7; and 87.68\% had abnormal INR values at lag 1 and 84.25\% at lag 7. These four laboratory tests (MPV, RDW, IgG, and INR) have high positive cross-correlations\footnote{Cross-correlation is used to measure the correlation between two time series at different time lags. The commonly used ``correlation" refers to cross-correlation at lag 0.} with the daily RBC demand at lags 1 and 7, as shown in Table \ref{crosscorrelation}. Thus, a higher number of patients with abnormal laboratory test results is associated with increased RBC demand, especially for MPV, RDW, IgG, and INR at lag 7. The clinical impact of these observations is being pursued in a separate study.

\begin{table}[htbp]
\small
\begin{tabular}{|l|c|c|}
\hline
\multirow{2}{*}{Abnormal laboratory test} & \multicolumn{2}{c|}{Cross-correlation} \\ \cline{2-3}
 & ~~~Lag 7~~~ & ~~~Lag 1~~~ \\ \hline
MPV & 0.52 & 0.45 \\ \hline
RDW & 0.52 & 0.48 \\ \hline
IgG & 0.53 & 0.52 \\ \hline
INR & 0.49 & 0.41 \\ \hline
\end{tabular}
\caption{Cross-correlation between daily RBC demand and abnormal laboratory test results at lag 7 and lag 1}
\label{crosscorrelation}
\end{table}

For semiweekly demand prediction, we directly calculate the results by aggregating the daily demand predictions on Tuesday to Thursday and on Friday to Monday. The mean (sd) of the predicted semiweekly demand on the test dataset is 325.00 (32.40) units, and the mean (sd) of the true daily demand is 323.50 (42.04) units. The RMSE is 39.22 and the MAPE is 8.97\% on the test dataset. We have also trained a separate hybrid model to predict semiweekly demand, however, the model performance (RMSE = 40.96 and MAPE = 9.96\%) is slightly worse than the results calculated based on daily predictions. Thus, for simplicity, we choose not to construct a separate model for the semiweekly demand prediction.

\subsection{Proposed ordering strategy}
\label{optimizationresult}

\begin{figure}[htbp]
\centering \makebox[\textwidth]{
\includegraphics[width=\textwidth]{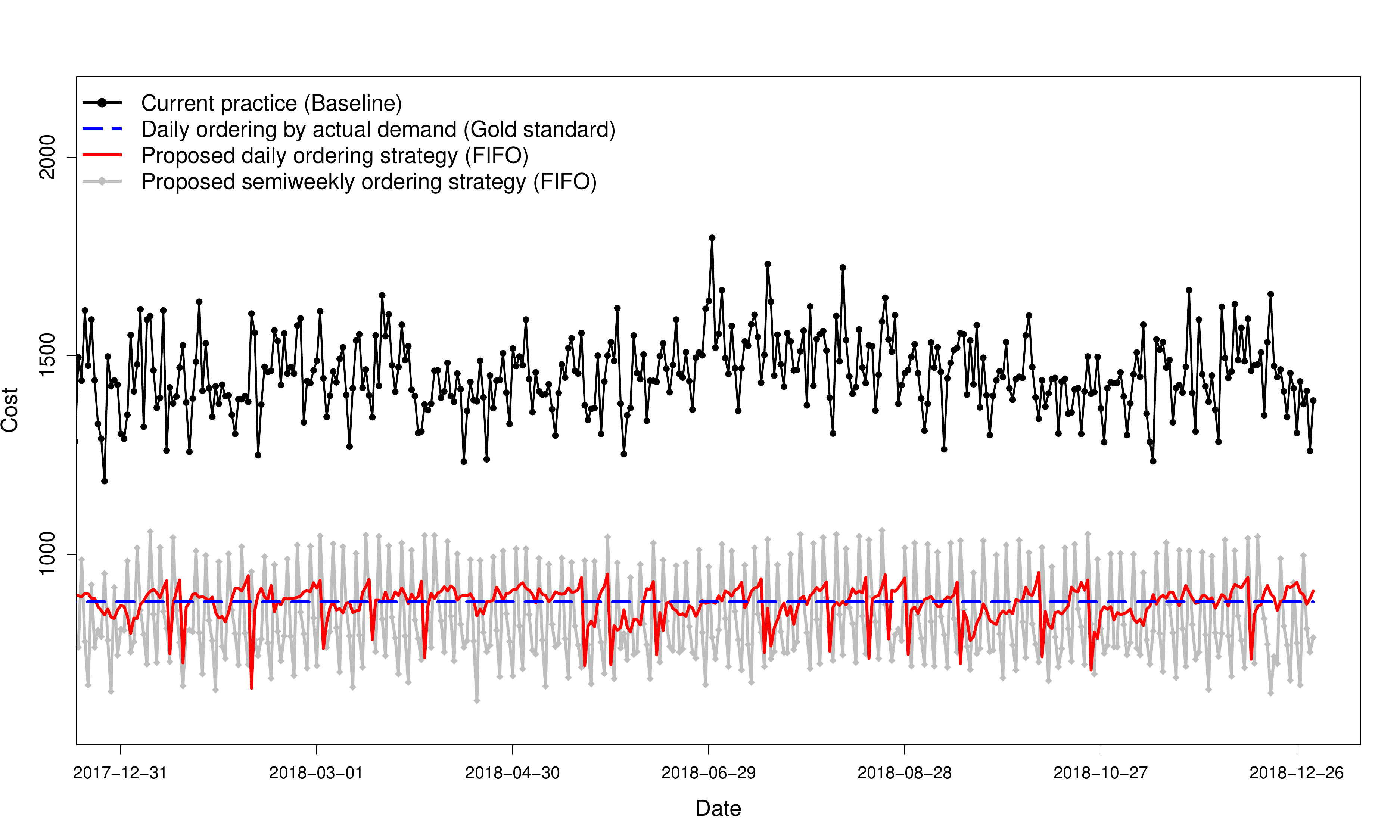}
}
\caption{Cost comparison of proposed ordering strategies versus baseline and gold standard}
\label{fig:costcomparison}
\end{figure}

The number of orders, inventory level, number of units requiring same-day urgent delivery, wastage, average cost, and total cost are calculated using Equations (\ref{obj1}) to (\ref{orderingpolicy}), assuming the routine delivery cost per order $a=100$, unit holding cost $h=1$, unit urgent delivery cost $u=300$, and unit wastage cost $w=50$. Using the procedure described in Section \ref{section:ordering}, the optimal inventory target is 1040 units, the optimal reorder level for daily orders is 830 units, and the optimal reorder level for semiweekly orders is 770 units. Comparisons are made during the time period of the test dataset (365 days) for four ordering strategies: current practice (baseline), ordering according to actual demand (gold standard), the proposed daily ordering strategy, and the proposed semiweekly ordering strategy. Ordering as current practice refers to the current ordering strategy used in hospital blood banks that reflects the actual order quantity, inventory level, and wastage. Since the actual number of units requiring urgent delivery was not captured in the database, the total cost calculated for the current ordering practice assumes that the urgent delivery cost is zero, and thus underestimates the actual cost. It is considered as the baseline strategy, whereas the results for ordering according to actual demand is considered as a gold standard. Table \ref{orderstrategycomparison} shows the results of the four ordering strategies. A summary of findings include:
\begin{enumerate}
 \item There is a 60.55\% reduction in the percentage of days with orders between the proposed semiweekly ordering strategy and baseline, whereas the proposed daily ordering strategy results in a 4\% reduction.
 \item The average order quantities for the proposed daily and semiweekly strategies over the entire period are less than the actual order quantities under current practice, and better reflect the actual demands. The mean order quantity for the semiweekly strategy on the 141 days with orders doubles the order quantity under current practice, which may raise operational problems such as requiring more packing boxes per order. Our collaborators indicate this is not of great concern for CBS.
 \item The average inventory level for the proposed daily strategy is significantly lower, a reduction of 41\%, as compared to the actual inventory level in hospital blood banks. Similarly, the proposed semiweekly ordering strategy results in a 39\% reduction of the inventory level. Both the means of the inventory levels for the daily and semiweekly strategies are close to the inventory level of the gold standard, but the semiweekly ordering strategy is associated with a larger variance since over 60\% of days have no orders. A leaner inventory leads to fresher RBC transfusions for patients. The DOH is reduced to 8.6 days for the proposed daily and semiweekly strategies from 12.47 days for current practice.
 \item There are no same-day urgent deliveries or wastage observed for the proposed daily and semiweekly strategies. However, there is no guarantee that this will always happen.
 \item The proposed daily and semiweekly ordering strategies achieve remarkable cost savings. Figure \ref{fig:costcomparison} illustrates the daily costs of the proposed daily strategy (red line), the semiweekly strategy (grey line), the actual costs under current practice (black line), and the constant cost of the gold standard (blue dashed line). Notably, the cost savings are driven by the significant decreases in the inventory level and routine order delivery costs. The proposed semiweekly strategy has the lowest average cost since it not only leads to a lower inventory level but also requires less frequent deliveries.
\end{enumerate}

Overall, both of the proposed daily and semiweekly ordering strategies result in a leaner inventory level, fresher blood, and lower costs, while no shortages (units requiring urgent delivery) or wastage are observed. Particularly, the semiweekly ordering strategy creates a ``win-win" situation, since it also provides a reduced, fixed delivery schedule that reduces costs and can free human resources both at CBS and hospital blood banks.

\begin{table}[htbp]
\footnotesize
\begin{tabular}{|p{0.22\textwidth}|p{0.16\textwidth}|p{0.18\textwidth}|p{0.15\textwidth}|p{0.15\textwidth}|}
\hline
\multirow{2}{*}{\textbf{Summary}} & \textbf{Current practice} & \textbf{Ordering according to actual   demand} & \textbf{Daily strategy} & \textbf{Semiweekly strategy} \\
 & \textbf{(Baseline)} & \textbf{(Gold standard)} & \textbf{(Proposed)} & \textbf{(Proposed)} \\ \hline
Number   (\%) of days with orders$\ddag$ & 362 (99.18\%) & 365 (100\%) & 347 (95.07\%) & 141 (38.63\%) \\ \hline
Order  quantity on days with orders - mean (sd)  & 106.02 (66.73) & 93.05 (27.01) & 97.98 (21.25) & 240.72 (133.73) \\ \hline
Daily   inventory level - mean (sd) & 1317.46 (65.89) & 780* (0) & 780.59 (33.68) & 800.98 (88.59) \\ \hline
Number   of units requiring urgent delivery - mean (sd) & N/A$\S$ & 0 & 0 & 0 \\ \hline
Number   of units wasted - mean (sd) & 0.75 (1.14) & 0 & 0 & 0 \\ \hline
Cost  - mean (sd) & 1454.03 (91.75) & 880 (0) & 875.66 (44.95) & 839.61 (115.65) \\ \hline
Total cost & 530722 & 321200 & 319617 & 306457   \\
(\% of baseline) &  & (60.52\%) & (60.22\%) & (57.74\%) \\ \hline
\end{tabular}
\caption{Comparisons among ordering strategies \\
\footnotesize{$\ddag$ The percentage is calculated by number of days with orders over 365 days of the test data. \\
* This number reflects the initial inventory level, $I_0$. \\
$\S$ The number of units requiring same-day urgent delivery was not available since the information was not captured in 2018.}}
\label{orderstrategycomparison}
\end{table}

\section{Discussion}
\label{discussion}
Improving the demand forecasting accuracy and building an efficient inventory management strategy for blood products are important to modern healthcare systems. It not only impacts the blood demand and supply management for blood suppliers and hospital blood banks, but may exert positive influence on patient outcomes. Among all fresh blood components, RBC is the most commonly administered therapeutic. It accounts for the largest portion of blood inventory, and determines the plan for blood collection and production. In Canada, due to the uncertain daily demand and the lack of quantitative evidence to support decision making, the RBC supply chain faces multiple challenges in hospital blood banks including excess inventory, highly variable ordering decisions, and over-frequent urgent orders. As a result, it has been very challenging for CBS to predict future demand and plan blood production.

In this study, we develop an STL + XGBoost hybrid algorithm for demand forecasting. It has the same level of prediction performance as a more complex LSTM model. It handles changing trend and seasonality, nonlinear patterns in residuals, and correlations among predictors in an efficient and accurate manner. We then construct a data-driven multi-period inventory problem for RBC ordering. The procedure to generate the data-driven ordering strategy involves solving for an optimal inventory target and an optimal reorder level through minimizing the absolute difference between the average costs using the predicted demands from the hybrid model and the actual demands. The proposed ordering strategy is a modified version of the classical $(s, S)$ policies considering a specific cost function. Shi et al. \cite{Shi2016} studied a nonparametric data-driven algorithm for the management of a stochastic periodic review inventory system with a constrained inventory target. There are two major differences between their optimization problem and ours: 1) The demand distribution is not stationary in our problem; 2) The cost function they considered does not include delivery costs, consequently there is no need to consider a reorder level to control the frequency of orders. They proved the asymptotical optimality of their algorithm under some technical assumptions and regularity conditions on the demand distribution. As a follow-up work, we plan to explore the optimality of our proposed ordering strategy in the setting for non-stationary demand. We expect that this will involve an appropriate asymptotic approach.

This study considers the aggregated RBC demand of all hospitals in Hamilton, Ontario for the development of the demand forecasting model rather than the demand of each hospital stratified by ABO blood groups. We chose to forecast the aggregate demand at the city level for the following reasons: i) All Hamilton hospital blood banks are managed by one Transfusion Medicine laboratory team. This is a common hospital blood bank management structure for Canadian cities, such as Toronto and Ottawa, Ontario. There are existing transaction networks available that allow blood delivery within cities at low costs. The central management structure enables the pooling of demand and inventory that may yield operational improvements. ii) The model accuracy is increased as the demand variability is decreased due to pooling. In other words, considering the aggregate demand can reduce the level of demand uncertainty. iii) The model provides important clinical predictors for the aggregated RBC demand of a diverse patient population in Hamilton, Ontario that could be representative of the overall Canadian RBC demand.

When considering ABO blood groups, there is no significant trend observed for each ABO blood group of transfused patients over the years. The proportions of ABO blood groups are consistent with the blood group distribution for Canadian population with very low variation. Our proposed methodology can be adapted to ensure sufficient inventory for each ABO blood group. Moreover, withdrawal policies to prioritize ABO identical RBC transfusion will be investigated in future studies.

When defining the RBC inventory optimization problem, the assumption of a fixed storage duration at CBS from the blood collection date to received date at hospital blood banks is a limitation of this work. Seasonality and nonlinear trend patterns were observed in the blood storage duration data from 2008 to 2018, reflecting the changes of the blood donation process at CBS. We found a significant increasing trend of the storage duration after 2017. This could be associated with new tools, such as chat bots and online appointment booking systems, launched at CBS in 2017 which increased the number of blood donations. Since the increase in blood supply exceeded the demand, a longer storage durations resulted at the CBS distribution centres. Both the storage duration at CBS and the DOH at hospital blood banks increased, consequently, the age of blood for transfusions has been increasing in the past couple of years. This also addresses the need of a better inventory management strategy that can help control the impacts of such changes at hospital blood banks, and share more accurate blood utilization information with CBS for blood collection planning.

To conclude, we propose a decision integration strategy for short-term demand forecasting and ordering for RBC blood components. It incorporates a robust and accurate hybrid model and a multi-period inventory optimization problem for ordering decisions. This leads to a significantly lower inventory level under a policy that has easy-to-compute order quantities and allows for a less frequent delivery schedule. It can potentially reduce the inventory by 40\% and decrease the number of deliveries by 60\%. The proposed ordering strategy can resolve the challenges faced by hospitals and CBS, and increase the transparency of blood utilization between blood suppliers and hospitals to promote an efficient blood supply chain, which may lead to better patient outcomes. Furthermore, the proposed data-driven ordering strategy is generalizable to other blood products or even other perishable products. We have initiated work to apply the proposed strategy to the inventory management of platelet components. We plan to develop a software application to implement the proposed methodology at hospital blood banks in Hamilton, Ontario in the near future, and expand to other hospital blood banks across Canada as a long-term plan.

\section{Acknowledgments}
This study was funded by Mitacs through the Accelerate Industrial Postdoc program (Grant Number: IT3639) in collaboration with Canadian Blood Services. The funding support from Canadian Blood Services was through the Blood Efficiency Accelerator program, funded by the federal government (Health Canada) and the provincial and territorial ministries of health. The views herein do not necessarily reflect the views of Canadian Blood Services or the federal, provincial, or territorial governments of Canada. The authors thank Dr. John Blake for his expertise in blood supply chain management. The authors thank Dr. Donald Arnold for providing valuable comments on the manuscript. The authors thank Tom Courtney, Rick Trifunov, Marianne Waito, and Masoud Nasari for arranging partner interaction activities, and providing information about blood collection, blood processing, and blood distribution at Canadian Blood Services. All final decisions regarding manuscript content were made by the authors.

\end{document}